\RequirePackage{fix-cm}
\documentclass[twocolumn,epjc3]{svjour3}  
\smartqed  
\usepackage{newtxtext,newtxmath} 
\RequirePackage{graphicx}
\RequirePackage{subfigure}
\RequirePackage{rotating}
\RequirePackage{pdflscape}
\RequirePackage{color}
\usepackage{mathtools}
\usepackage{extarrows}
\usepackage{bm}
\RequirePackage{latexsym}
\RequirePackage{xspace}
\RequirePackage[numbers,sort&compress]{natbib}
\RequirePackage[colorlinks,citecolor=blue,urlcolor=blue,linkcolor=blue]{hyperref}
\usepackage{amsmath,amsfonts,bm}
\usepackage{wasysym}
\usepackage{color}
\usepackage{lineno}
\usepackage[usenames,dvipsnames,svgnames,table]{xcolor}	
\let\oldequation\equation
\let\oldendequation\endequation

\renewenvironment{equation}
  {\linenomathNonumbers\oldequation}
  {\oldendequation\endlinenomath}

\let\oldalign\align
\let\oldendalign\endalign

\renewenvironment{align}
  {\linenomathNonumbers\oldalign}
  {\oldendalign\endlinenomath}

\newcommand{\vbb}{$\nu\beta\beta$\xspace}
\newcommand{\thalf}{$T_{1/2}^{0\nu}$}

\newcommand{\Q}{CUORE\xspace}

\newcommand{\cumo}{CUPID-Mo\xspace}
\newcommand{\kgyr}{kg$\times$year\xspace}

\newcommand{\qbb}{$Q_{\beta\beta}$\xspace}

\newcommand{\lmo}{Li$_2$MoO$_4$\xspace}
\newcommand{\enrLMO}{Li$_{2}${}$^{100}$MoO$_4$\xspace}
\newcommand{\isomo}{$^{100}$Mo\xspace}

\newcommand{\mone}{$\mathcal{M}_{1}$\xspace}
\newcommand{\mtwo}{$\mathcal{M}_{2}$\xspace}

\newcommand{\deltat}{$\Delta t$\xspace}
\newcommand{\betagamma}{$\beta/\gamma$\xspace}
\newcommand{\exposuresymb}{Mt\xspace}

\newcommand{\Co}{$^{60}\mathrm{Co}$\xspace}

\newcommand{\ky}{kg$\times$year\xspace}
\newcommand{\kd}{kg$\times$day\xspace}
\newcommand{\ckky}{counts/(keV$\times$kg$\times$year)\xspace}

\def\exposure{\updateValue{2.71}\xspace}
\def\isoexposure{\updateValue{1.47}\xspace}

\def\analysiseff{\updateValue{88.4 $\pm$ 1.8}\xspace}

\def\roiMedian{\updateValue{17.12~\,keV}\xspace}
\def\roiWidth{\updateValue{$(17.1 \pm 4.5)$}~\,keV\xspace}

\def\totaleff{\updateValue{$67.1 \pm 1.7$}\xspace}

\def\scalefactor{\updateValue{$1.126 \pm 0.052$}\xspace}
\def\resolution{\updateValue{$(7.4 \pm 0.4)~\,\mathrm{keV}$}\xspace}
\def\ebias{\updateValue{$(-0.42 \pm 0.30)~\,\mathrm{keV}$}\xspace}

\def\rateNumber{\updateValue{$3.8 \times 10^{-25} \ \mathrm{y^{-1}}$}\xspace}

\def\rate{\updateValue{\Gamma^{0\nu} < 3.8 \times 10^{-25} \ \mathrm{year^{-1}}  \ \text{(stat.+syst.}) \ \text{at 90\% C.I.} \xspace}}

\def\ckky{$\mathrm{counts/(keV\cdot kg \cdot year)}$\xspace}

\def\posteriorexp{\updateValue{\lambda = (6.061 \pm 0.001)\times 10^{24} \ \mathrm{year}}\xspace}

\newcommand{\updateValue}[1]{{\color{Black}{#1}}}

\def\thalf{$T_{1/2}^{0\nu}$\xspace}
\def\hlife{\updateValue{1.8}}
\newcommand{\hlifeResult}{\thalf $> \hlife \times10^{24}$~year\ \text{(stat.+syst.)} \ \text{at 90\% CI.}\xspace}

\newcommand{\hlifeResultNEMO}{\thalf $> 1.1 \times10^{24}$~year \ \text{at 90\% C.I.}\xspace}
\newcommand{\hlifeResultAMORE}{\thalf $> 9.5 \times10^{22}$~year \ \text{at 90\% C.I.}\xspace}

\newcommand{\hlifeResultEqn}{T_{1/2}^{0\nu} > \hlife \times10^{24}~\text{year} \ \text{(stat.+syst.)} \ \text{at 90\% CI.}\xspace}

\def\mbbrange{\updateValue{$(0.28$--$0.49)$}\xspace}
\def\mbbavg{$\left<m_{\beta \beta}\right>$\xspace}
\newcommand{\mbbResult}{\mbbavg $<$ \mbbrange~\,\text{eV}\xspace}
\newcommand{\mbbResultEqn}{\mbox{\mbbavg $<$ \mbbrange~\,eV}\xspace}

\def\mbbrangeNEMO{\updateValue{$(0.33$--$0.62)$}\xspace}
\newcommand{\mbbResultNEMO}{\mbbavg $<$ \mbbrangeNEMO~\,\text{eV}\xspace}

\def\mbbrangeAMORE{\updateValue{$(1.2$--$2.1)$}\xspace}
\newcommand{\mbbResultAMORE}{\mbbavg $<$ \mbbrangeAMORE~\,\text{eV}\xspace}

\def\bgidx{\updateValue{$\left(4.7\pm 1.7\right) \times 10^{-3}$}\xspace}
\newcommand{\bi}{\mbox{$B$ = \bgidx~\,\ckky}\xspace}

\def\bgidxpost{\updateValue{$\left(3.9^{+1.7}_{-1.6} \right) \times 10^{-3}$}\xspace}
\newcommand{\bipost}{\mbox{$B$ = \bgidxpost~\,\ckky}\xspace}

\journalname{Eur. Phys. J. C}
\begin{document}

\title{Final results on the $0\nu\beta\beta$ decay half-life limit of $^{100}$Mo from the CUPID-Mo experiment}


\author{
C.~Augier\thanksref{IPNL}\and
A.~S.~Barabash\thanksref{ITEP}\and
F.~Bellini\thanksref{Sapienza,INFN-Roma}\and
G.~Benato\thanksref{CEA-IRFU,LNGS}\and
M.~Beretta\thanksref{UCB}\and
L.~Berg\'e\thanksref{IJCLab}\and
J.~Billard\thanksref{IPNL}\and
Yu.~A.~Borovlev\thanksref{NIIC}\and
L.~Cardani\thanksref{INFN-Roma}\and
N.~Casali\thanksref{INFN-Roma}\and
A.~Cazes\thanksref{IPNL}\and
M.~Chapellier\thanksref{IJCLab}\and
D.Chiesa\thanksref{Milano,INFN-Milano}\and
I.~Dafinei\thanksref{INFN-Roma}\and
F.~A.~Danevich\thanksref{KINR}\and
M.~De~Jesus\thanksref{IPNL}\and
P.~de~Marcillac\thanksref{IJCLab}\and
T.~Dixon\thanksref{IJCLab}\and
L.~Dumoulin\thanksref{IJCLab}\and
K.~Eitel\thanksref{KIT-IK}\and
F.~Ferri\thanksref{CEA-IRFU}\and
B.~K.~Fujikawa\thanksref{LBNLNSD}\and
J.~Gascon\thanksref{IPNL}\and
L.~Gironi\thanksref{Milano,INFN-Milano}\and
A.~Giuliani\thanksref{e1,IJCLab}\and
V.~D.~Grigorieva\thanksref{NIIC}\and
M.~Gros\thanksref{CEA-IRFU}\and
D.~L.~Helis\thanksref{CEA-IRFU,GSSI}\and
H.~Z.~Huang\thanksref{Fudan}\and
R.~Huang\thanksref{UCB}\and
L.~Imbert\thanksref{IJCLab}\and
J.~Johnston\thanksref{MIT}\and
A.~Juillard\thanksref{IPNL}\and
H.~Khalife\thanksref{CEA-IRFU}\and 
M.~Kleifges\thanksref{KIT-IPE}\and
V.~V.~Kobychev\thanksref{KINR}\and
Yu.~G.~Kolomensky\thanksref{UCB,LBNLNSD}\and
S.I.~Konovalov\thanksref{ITEP}\and
P.~Loaiza\thanksref{IJCLab}\and
L.~Ma\thanksref{Fudan}\and
E.~P.~Makarov\thanksref{NIIC}\and
R.~Mariam\thanksref{IJCLab}\and
L.~Marini\thanksref{UCB,GSSI}\and
S.~Marnieros\thanksref{IJCLab}\and
X.-F.~Navick\thanksref{CEA-IRFU}\and
C.~Nones\thanksref{CEA-IRFU}\and
E.B.~Norman\thanksref{UCBNE}\and
E.~Olivieri\thanksref{IJCLab}\and
J.~L.~Ouellet\thanksref{MIT}\and
L.~Pagnanini\thanksref{LNGS}\and
L.~Pattavina\thanksref{LNGS,TUM}\and
B.~Paul\thanksref{CEA-IRFU}\and
M.~Pavan\thanksref{Milano,INFN-Milano}\and
H.~Peng\thanksref{USTC}\and
G.~Pessina\thanksref{INFN-Milano}\and
S.~Pirro\thanksref{LNGS}\and
D.~V.~Poda\thanksref{IJCLab}\and
O.~G.~Polischuk\thanksref{KINR}\and
S.~Pozzi\thanksref{INFN-Milano}\and
E.~Previtali\thanksref{Milano,INFN-Milano}\and
Th.~Redon\thanksref{IJCLab}\and
A.~Rojas\thanksref{LSM}\and
S.~Rozov\thanksref{JINR}\and 
V.~Sanglard\thanksref{IPNL}\and
J.A.~Scarpaci\thanksref{IJCLab}\and
B.~Schmidt\thanksref{UCB,NW}\and
Y.~Shen\thanksref{Fudan}\and
V.~N.~Shlegel\thanksref{NIIC}\and
V.~Singh\thanksref{UCB}\and
C.~Tomei\thanksref{INFN-Roma}\and
V.~I.~Tretyak\thanksref{KINR}\and 
V.~I.~Umatov\thanksref{ITEP}\and
L.~Vagneron\thanksref{IPNL}\and
M.~Vel\'azquez\thanksref{UGA}\and
B.~Welliver\thanksref{UCB}\and
L.~Winslow\thanksref{MIT}\and
M.~Xue\thanksref{USTC}\and
E.~Yakushev\thanksref{JINR}\and
M.~Zarytskyy\thanksref{KINR}\and
A.~S.~Zolotarova\thanksref{CEA-IRFU}
}

\thankstext{e1}{e-mail: andrea.giuliani@ijclab.in2p3.fr}

\institute{
Univ Lyon, Universit\'{e} Lyon 1, CNRS/IN2P3, IP2I-Lyon, F-69622, Villeurbanne, France  \label{IPNL} \and
National Research Centre Kurchatov Institute, Institute of Theoretical and Experimental Physics, 117218 Moscow, Russia  \label{ITEP} \and 
Dipartimento di Fisica, Sapienza Universit\`a di Roma, P.le Aldo Moro 2, I-00185, Rome, Italy \label{Sapienza} \and 
INFN, Sezione di Roma, P.le Aldo Moro 2, I-00185, Rome, Italy \label{INFN-Roma} \and
IRFU, CEA, Universit\'{e} Paris-Saclay, F-91191 Gif-sur-Yvette, France  \label{CEA-IRFU} \and 
INFN, Laboratori Nazionali del Gran Sasso, I-67100 Assergi (AQ), Italy \label{LNGS} \and
Department of Physics, University of California, Berkeley, California 94720, USA \label{UCB} \and
Université Paris-Saclay, CNRS/IN2P3, IJCLab, 91405 Orsay, France \label{IJCLab} \and
Nikolaev Institute of Inorganic Chemistry, 630090 Novosibirsk, Russia \label{NIIC} \and
Dipartimento di Fisica, Universit\`{a} di Milano-Bicocca, I-20126 Milano, Italy \label{Milano} \and 
INFN, Sezione di Milano-Bicocca, I-20126 Milano, Italy \label{INFN-Milano} \and 
Institute for Nuclear Research, 03028 Kyiv, Ukraine \label{KINR} \and 
Institute for Astroparticle Physics, Karlsruhe Institute of Technology, 76021 Karlsruhe, Germany \label{KIT-IK} \and
Nuclear Science Division, Lawrence Berkeley National Laboratory, Berkeley, California 94720, USA \label{LBNLNSD} \and
Gran Sasso Science Institute, L’Aquila I-67100, Italy \label{GSSI} \and
Key Laboratory of Nuclear Physics and Ion-beam Application (MOE), Fudan University, Shanghai 200433, PR China \label{Fudan} \and
Massachusetts Institute of Technology, Cambridge, MA 02139, USA \label{MIT} \and
Institute for Data Processing and Electronics, Karlsruhe Institute of Technology, 76021 Karlsruhe, Germany \label{KIT-IPE} \and
Department of Nuclear Engineering, University of California, Berkeley, California 94720, USA \label{UCBNE} \and
Physik Department, Technische Universit\"at M\"unchen, Garching D-85748, Germany \label{TUM} \and
Department of Modern Physics, University of Science and Technology of China, Hefei 230027, PR China \label{USTC} \and
LSM, Laboratoire Souterrain de Modane, 73500 Modane, France \label{LSM} \and
Laboratory of Nuclear Problems, JINR, 141980 Dubna, Moscow region, Russia \label{JINR} \and
Department of Physics \& Astronomy, Northwestern University, Evanston, Illinois 60208, USA \label{NW} \and
Universit\'e Grenoble Alpes, CNRS, Grenoble INP, SIMAP, 38402 Saint Martin d'H\'eres, France \label{UGA}
}

\date{Received: date / Accepted: date}

\maketitle


\begin{abstract}
    The CUPID-Mo experiment to search for 0\vbb decay in \isomo has been recently completed after about 1.5 years of operation at Laboratoire Souterrain de Modane (France). It served as a demonstrator for CUPID, a next generation 0\vbb decay experiment. CUPID-Mo was comprised of 20 enriched \enrLMO scintillating calorimeters, each with a mass of $\sim$ 0.2~\,kg, operated at $\sim$20~\,mK. We present here the final analysis with the full exposure of CUPID-Mo (\isomo exposure of \isoexposure \kgyr) used to search for lepton number violation via 0\vbb decay. We report on various analysis improvements since the previous result on a subset of data, reprocessing all data with these new techniques. We observe zero events in the region of interest and set a new limit on the \isomo 0\vbb decay half-life of \hlifeResult Under the light Majorana neutrino exchange mechanism this corresponds to an effective Majorana neutrino mass of \mbbResult, dependent upon the nuclear matrix element utilized.
\end{abstract}


\section{Introduction}
\label{sec:intro}

Ever since the neutrino was found to have mass via the observation of flavor state oscillations~\cite{ PhysRevLett.81.1562, Ahmad2001}, the nature of the neutrino mass itself has remained a mystery. Unlike charged leptons, neutrinos may be Majorana \cite{MajoranaPaper, PhysRevD.22.2227} instead of Dirac particles. In addition to implying that neutrinos and anti-neutrinos would be the same particle~\cite{Racah1938}, this would also imply that the total lepton number is not conserved~\cite{Pontecorvo1968}. This may provide for a possible explanation of baryon asymmetry (i.e., the imbalance between matter and anti-matter) in the early universe~\cite{FUKUGITA198645, DAVIDSON2008105}. 

Two-neutrino double-beta (2\vbb) decay is a rare Standard Model process which can occur in some even-even nuclei for which single beta decays are energetically forbidden (or heavily disfavored due to large changes in angular momentum). In this process two neutrons are converted into two protons with the emission of two electrons and two electron anti-neutrinos. The possibility of the neutrino being a Majorana fermion raises the prospect that neutrinoless double-beta (0\vbb) decay may occur~\cite{Furry1939, PhysRevD.25.2951}. In the case of 0\vbb decay, we would again observe the conversion of two neutrons into two protons, but such a decay would only produce two electrons. Whereas 2\vbb decay conserves lepton number, 0\vbb decay would result in an overall violation of the total lepton number by two units~\cite{Bilenky2015, Goswami2015, Dolinski2019}, and indicate physics beyond the Standard Model. At present this process is unobserved with limits on the decay half-life at the level of $10^{24}-10^{26}$~\,year via the isotopes $^{76}$Ge, $^{82}$Se, \isomo, $^{130}$Te, and $^{136}$Xe~\cite{GERDA2020, MJD2019, CUPID0Final,CUORE1ton, CUOREPRL2020, CuMoPRL, NEMO30vMo, KamLANDZen2016, EXO200.2019}. The process of 0\vbb decay has several possible mechanisms~\cite{Bilenky2015, Dolinski2019, Deppisch2012, Rodejohann2012, PhysRevD.68.034016, Atre2009, Blennow2010, MITRA201226, Cirigliano2018}, however the minimal extension to the Standard Model provides the simplest via the exchange of a light Majorana neutrino. The rate of this process is dependent upon the square of the effective Majorana neutrino mass, \mbbavg:
\begin{equation}
    \Gamma^{0\nu}= G^{0\nu}g^{4}_{A}|M^{0\nu}_{\beta\beta}|^2|\langle m_{\beta\beta} \rangle |^2/m_e^2,
\end{equation}
where $g_{A}$ is the weak axial vector coupling constant, $M^{0\nu}_{\beta\beta}$ is the nuclear matrix element, $G_{0\nu}$ is the decay phase space, and $m_e$ is the electron mass. The effective mass is a linear combination of the three neutrino mass eigenstates, with the present limits on \mbbavg ranging from 60--600~\,meV~\cite{Dolinski2019}.

A vibrant experimental field has emerged to search for 0\vbb decay with experiments using a variety of nuclei and a wide range of methods (see~\cite{Dolinski2019} for a review of some of these). The main experimental signature for this decay is a peak of the summed electron energy at the $Q$-value of the decay (\qbb; the difference in energy between the parent and daughter nuclei) broadened only by detector energy resolution. There are 35 natural $\beta\beta$ decay isotopes~\cite{TRETYAK200283}, however from an experimental perspective only a subset of these are relevant. Searching for 0\vbb decay requires that the number of target atoms should be very large and that the background rate should be small. In an ideal case a candidate isotope should have \qbb > 2.6~\,MeV so it is above significant natural $\gamma$ backgrounds and has the phase space for a relatively fast decay rate. It should also occur with a high natural abundance or be easily enriched.

The isotope \isomo meets these requirements with a \qbb of 3034~\,keV and a relatively favorable phase space compared to other isotopes. Additionally it is relatively easy to enrich detectors with \isomo. Several experiments have utilized \isomo for 0\vbb decay searches: NEMO-3, LUMINEU, and AMoRE. NEMO-3 utilized foils containing \isomo and used external sensors to measure time of flight and provide calorimetry. It ran from 2003-2010 at Modane and accumulated an exposure of 34.3 \ky in \isomo. It found no evidence for 0\vbb and set a limit of \hlifeResultNEMO with a limit on the effective Majorana mass of \mbbResultNEMO~\cite{NEMO30vMo}. LUMINEU was a pilot experiment for \isomo based calorimeters and served as a precursor to CUPID-Mo. LUMINEU utilized both Zn$^{100}$MoO$_{4}$ and \enrLMO crystals, and found \enrLMO is more favorable for 0\vbb decay searches~\cite{lumineu2017}. AMoRE operates at Yangyang underground laboratory in Korea. The AMoRE-pilot operated with $^{48\text{depl}}$Ca$^{100}$MoO$_{4}$ crystals. As with CUPID-Mo, AMoRE utilizes light and heat to provide particle identification. The AMoRE-pilot has set a limit using 111 \kd of exposure of \hlifeResultAMORE with a corresponding effective Majorana mass limit of \mbbResultAMORE~\cite{AMORE2019}.

Scintillating calorimeters are one of the most promising current technologies for 0\vbb decay searches, with many possible configurations~\cite{DPoda2021, PirroScintBolo, Tabarelli2009, Cardani2013LMO, Giuliani2018, CUPID0Final, AMORE2019,sym13122255}. These consist of a crystalline material, containing the source isotope, capable of scintillating at low temperatures which is operated as a cryogenic calorimeter coupled to light detectors to detect scintillation light. Particle identification is based on the difference in scintillation light produced for a given amount of energy deposited in the main calorimeter. This technology has demonstrated excellent energy resolution, high detection efficiency, and low background rates (due to the rejection of $\alpha$ events). The rejection of $\alpha$ events is of primary concern as the energy region above $\sim$2.6~\,MeV is populated by surface radioactive contaminants with degraded energy collection of $\alpha$ particles~\cite{CUORE1ton, CUOREPRL2020}.

CUPID (CUORE Upgrade with Particle IDentification) is a next-generation experiment \cite{CUPID:2019} which will use this scintillation calorimeter technology. It will build on the success of CUORE (Cryogenic Underground Observatory for Rare Events) which demonstrated the feasibility of a tonne-scale experiment using cryogenic calorimeters \cite{CUORE_cryostat,CUORE1ton}. In this paper we describe the final 0\vbb decay search results of the CUPID-Mo experiment which has successfully demonstrated the use of $^{100}$Mo-enriched \lmo detectors for CUPID. In sections \ref{sec:cumo} and \ref{sec:data} we introduce the CUPID-Mo experiment and an overview of the collected data. In sections \ref{sec:dataprod} and \ref{sec:additionalProd} we describe the data production and basic data quality selection. Then in sections \ref{sec:psd}--\ref{sec:other} we describe in detail the improved data selection cuts we use to reduce experimental background rates. We then describe our Bayesian 0\vbb decay analysis in sections \ref{sec:eff}--\ref{sec:bayes}. Finally, the results and their implications are discussed in section \ref{sec:results}. 
\section{CUPID-Mo Experiment}
\label{sec:cumo}
\begin{figure}[htpb]
    \centering
    \includegraphics[width=\columnwidth]{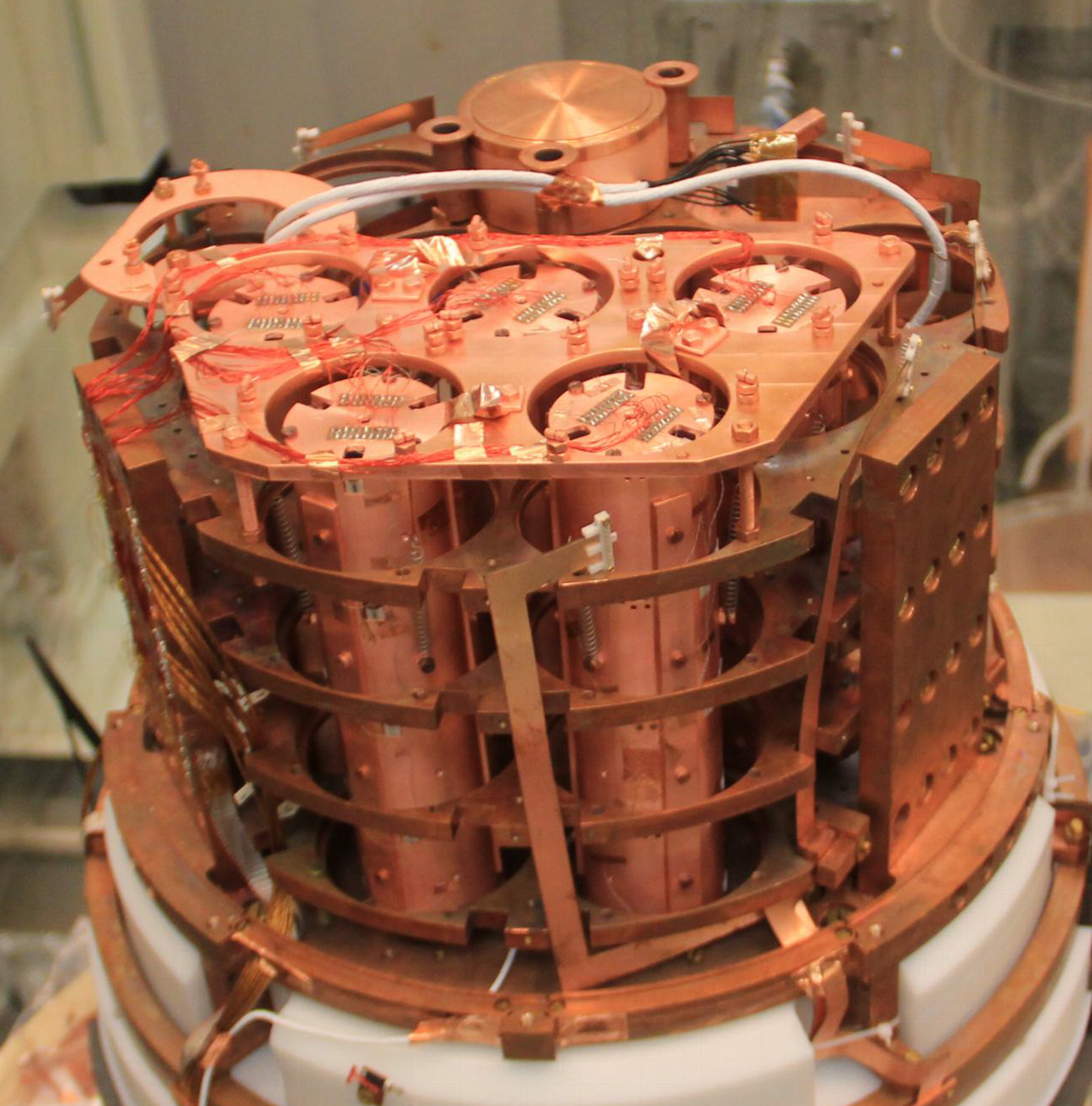}
    \includegraphics[width=\columnwidth]{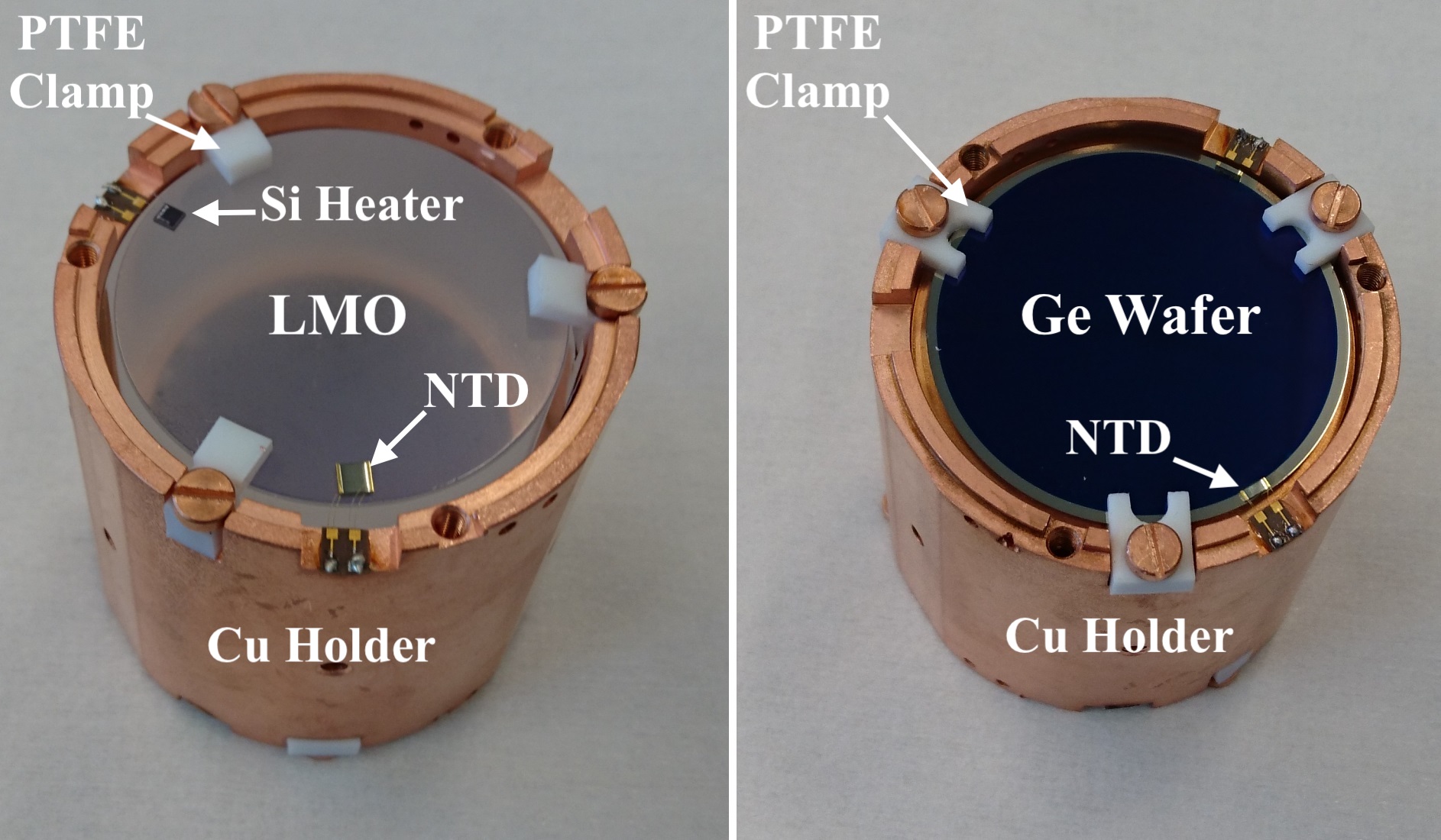}
    \caption{Images showing the CUPID-Mo detector array (5 nearest towers) mounted in the EDELWEISS cryostat (\textit{Top}) and a single module assembled in the Cu holder (\textit{Bottom})~\cite{CuMo_instrument}. (\textit{Bottom left})  view from the top on the LMO detector, NTD-Ge, Si heater, copper holder and PTFE clamps. (\textit{Bottom right})  view from the bottom on the Ge LD and its NTD-Ge thermistor and PTFE clamps. }
    \label{fig:exp}
\end{figure}

The \cumo experiment was operated underground at the Laboratoire Souterrain
de Modane in France~\cite{CuMo_instrument} following a successful pilot experiment, LUMINEU~\cite{lumineu2017, Poda2017}. The \cumo detector array was comprised of 20 scintillating \lmo (LMO) cylindrical crystals, $\sim$210~\,g each (see Fig.~\ref{fig:exp}). These are enriched in \isomo to $\sim$97\%, and operated as cryogenic calorimeters at $\sim20$~\,mK. Each LMO detector is paired with a Ge wafer light detector (LD) and assembled into a detector module with a copper holder and Vikuiti$^{\text{TM}}$ reflective foil to increase scintillation light collection. Both the LMO detectors and LDs are instrumented with a neutron-transmutation doped Ge-thermistor (NTD)~\cite{Haller1984} for data readout. Additionally, a Si heater is attached to each LMO crystal which is used to monitor detector performance.


The modules are organized into five towers with four floors and mounted in the EDELWEISS cryostat~\cite{Armengaud_2017} (see Fig.~\ref{fig:exp}). In this configuration each LMO detector (apart from those on the top floor) nominally has two LDs increasing the discrimination power. We note that one LD did not function resulting in two LMO detectors which are not on the top floor having only a single working LD.

CUPID-Mo has demonstrated excellent performance, crystal radiopurity, energy resolution, and high detection efficiency~\cite{CuMo_instrument}, close to the requirements of the CUPID experiment~\cite{CUPID:2019}. An analysis of the initial CUPID-Mo data (1.17 \ky of $^{100}$Mo exposure) led to a limit on the half-life of 0\vbb decay in $^{100}$Mo of $T_{1/2}^{0\nu}>1.5\times 10^{24}$yr~ at 90\% C.I. \cite{CuMoPRL}. For the final results of CUPID-Mo we increase the exposure and also develop novel analysis procedures which will be critical to allow CUPID to reach its goals.
\section{CUPID-Mo Data Taking}
\label{sec:data}
The data utilized in this analysis was acquired from early 2019 through mid-2020 (481 days in total) with a duty cycle of $\sim$89\% of the EDELWEISS cryogenic facility. The data collected between periods of cryostat maintenance or special calibrations, which require the external shield to open, are grouped into ``datasets'' typically $\sim$1--2 months long. Within each dataset we attempt to have periods of calibration data taking (typically, $\sim$2-d-long measurements every $\sim$10 days) bracketing physics data taking, corresponding to 21\% and 70\% of the total CUPID-Mo data respectively. \cumo utilizes a U/Th source placed outside the copper screens of the cryostat (see~\cite{CuMo_instrument}) for standard LMO detector calibration, providing a prominent $\gamma$ peak at 2615~\,keV, as well as several other peaks at lower energies to perform calibration. The primary calibration source is a thorite mineral with $\sim$50\,Bq of $^{232}$Th, and $\sim$100\,Bq of $^{238}$U with significantly smaller activity from $^{235}$U. Overall, nine datasets are utilized in this final analysis with a total LMO exposure of \exposure~\,\kgyr, corresponding to a \isomo exposure of \isoexposure~\,\kgyr.  As was the case in the previous analysis~\cite{CuMoPRL}, we exclude three short periods of data taking which have an insufficient amount of calibration data to adequately perform thermal gain correction, and determine the energy calibration. We also exclude one LMO detector which has abnormally poor performance in all datasets.

Additional periods of data taking with a very high activity $^{60}$Co source ($\sim$100\,kBq, $\sim$2\% of CUPID-Mo data) were performed near regular liquid He refills (every $\sim$ 10 days). While the \Co source was primarily used for EDELWEISS~\cite{Armengaud_2017}, it was also utilized in \cumo for LD calibration via X-ray fluorescence~\cite{CuMo_instrument} and is further described in section~\ref{sec:ldcalib}. The remainder of the data in \cumo is split between calibration with a $^{241}$Am+$^{9}$Be neutron source (2\%) and a $^{56}$Co calibration source ($\sim$5\%).



\section{Data Production}
\label{sec:dataprod}

\begin{figure}[htpb]
    \centering
    \includegraphics[width=\columnwidth]{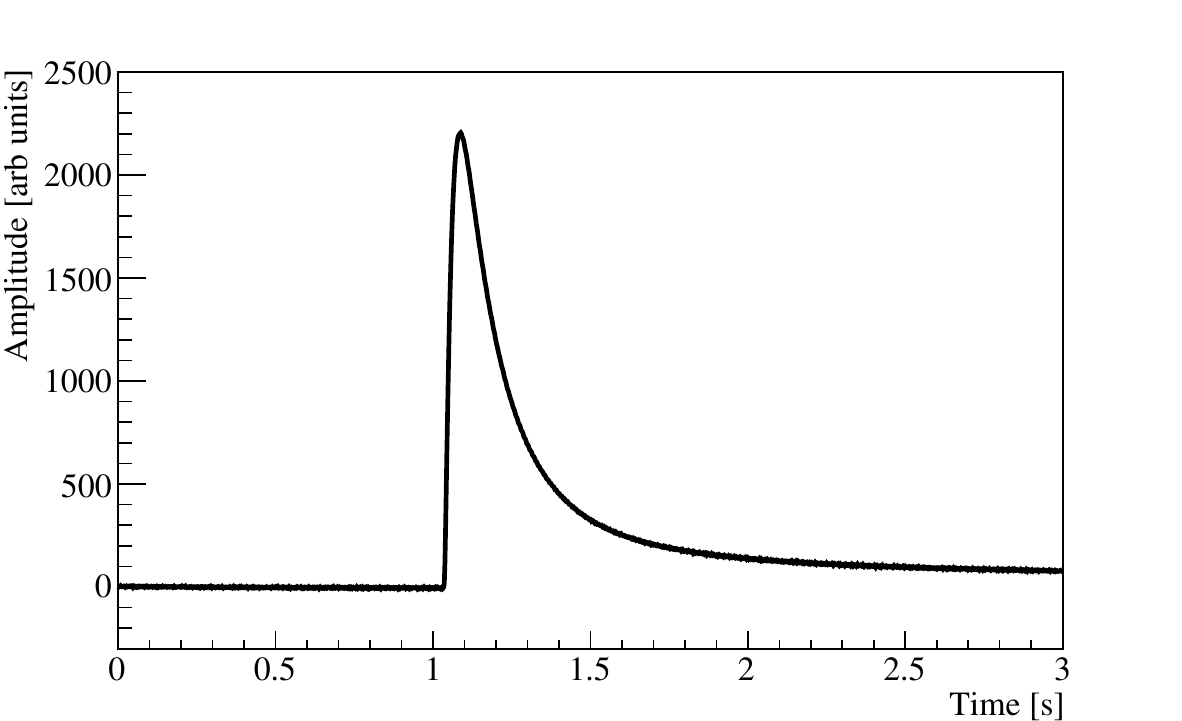}
    \includegraphics[width=\columnwidth]{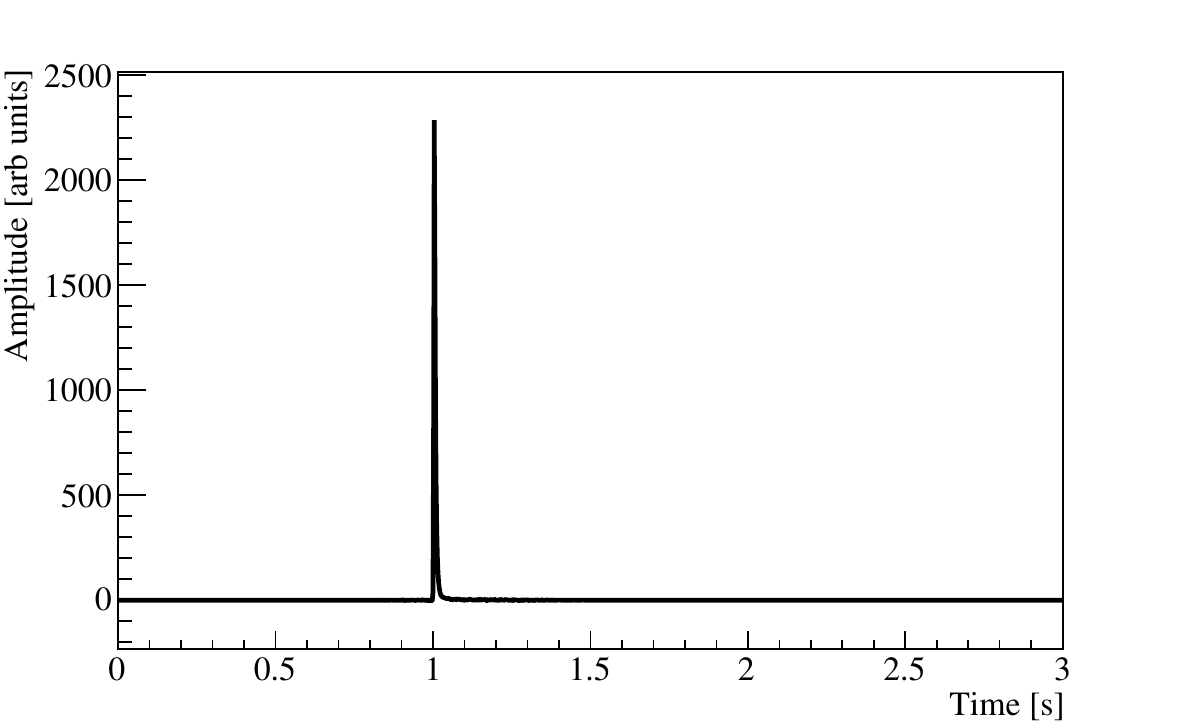}
    \caption{Typical average pulses for LMO detector (top) and LD (bottom) readout. Note that the LMO pulses are significantly longer in duration owing to the larger heat capacity of the LMO compared to the much smaller Ge LD.}
    \label{fig:ap}
\end{figure}

We outline here the basic data production steps required to create a calibrated energy spectrum. Starting with AC biased NTDs, we perform demodulation in hardware and sample the resulting voltage signals from all heat and light channels at 500~\,Hz to produce the raw data. We then utilize the Diana and Apollo framework~\cite{PhysRevC.93.045503, Apollo_Domizio_2018}, developed by the \Q-0, \Q, and CUPID-0 collaborations, with modifications for \cumo. Events in data are triggered ``offline'' in Apollo using the optimum trigger method~\cite{OT_Domizio_2011} to search for pulses. This method requires an initial triggering of the data to construct an average pulse template and average noise power spectrum. This in turn is used to build an optimum filter (OF) which maximizes the signal-to-noise ratio. This OF is then used as the basis for the primary triggering. An event is triggered when the filtered data crosses a set threshold relative to the typical OF resolution obtained from the average noise power spectrum for a given channel (set at a value of 10$\sigma$). We periodically inject flags to indicate noise triggers into the data stream in order to obtain a sample of noise events which allows us to characterize the noise on each channel. For this data production we utilize a 3 s time window for both the heat and light channels. This is long enough to allow sufficient time for the LMO waveform to return towards baseline whilst being short enough to keep the rate of pileup events relatively low. This choice also keeps the event windows of equal size between the LMO detectors and LDs (see Fig.~\ref{fig:ap}). The first 1 s of data prior to the trigger is the pretrigger window which is used in pulse baseline measurements. For reference the typical 10\%--90\% rise and 90\%--30\% fall times for the LMO detectors are $\sim$20~\,ms and $\sim$300~\,ms respectively, and for the LDs they are much shorter at $\sim$4~\,ms and $\sim$9~\,ms respectively~\cite{CuMo_instrument}.

Once triggered data is available, basic event reconstruction quantities are computed, such as the waveform average baseline (the mean of the waveform in the first 80\% of the pretrigger window), baseline slope, pulse rise and decay times, and other parameters that are computed directly on the raw waveform. A mapping of so-called ``side'' channels is generated, grouping the LDs that a given LMO crystal directly faces in the data processing framework. In each dataset, a new OF is constructed for each channel, and used to estimate the amplitude of both the LMO detector and LD events, the latter being restricted to search in a narrow range around the LMO event trigger time. After the OF amplitudes are available, thermal gain correction is performed on the LMO detectors (see section~\ref{sec:stab}) and finally the LMO detector energy scale is calibrated from the external U/Th calibration runs (see section~\ref{sec:calib}). Each step of the data production is done on runs within a single dataset, with the exception of the first two datasets which share a common thermal gain correction and energy calibration period to boost statistics.
\subsection{Thermal Gain Correction}
\label{sec:stab}
After we have reconstructed pulse amplitudes via the OF we must perform a thermal gain correction (sometimes referred to as ``stabilization'')~\cite{Alessandrello1998}. This process corrects for thermal-gain changes in detector response which cause slight differences in pulse amplitude for a given incident energy, resulting in artificially broadened peaks.
The pulse baseline is used as a proxy for the temperature, allowing us to use it to correct for thermal-gain changes due to fluctuations in temperature. This correction uses calibration data, from which we select a sample of events determined to be the 2615~\,keV $\gamma$-ray full absorption peak from $^{208}$Tl. We perform a fit of the OF amplitudes ($A$) as a function of the mean baselines ($b_{\mu}$) given by the linear function $f(b_{\mu})= p_0+p_1\cdot b_{\mu}$ and compute the scaled corrected amplitude ($\tilde{C}$) as $\tilde{C} = (A/f(b_{\mu}))\cdot 2615$. This correction is applied to both calibration and physics data within a dataset. We observe that the LDs do not demonstrate any significant thermal gain drift and as such do not perform this step on them.
\subsection{LMO Detector Calibration}
\label{sec:calib}
\begin{figure*}[htbp]
    \centering
    \includegraphics[width=0.49\textwidth]{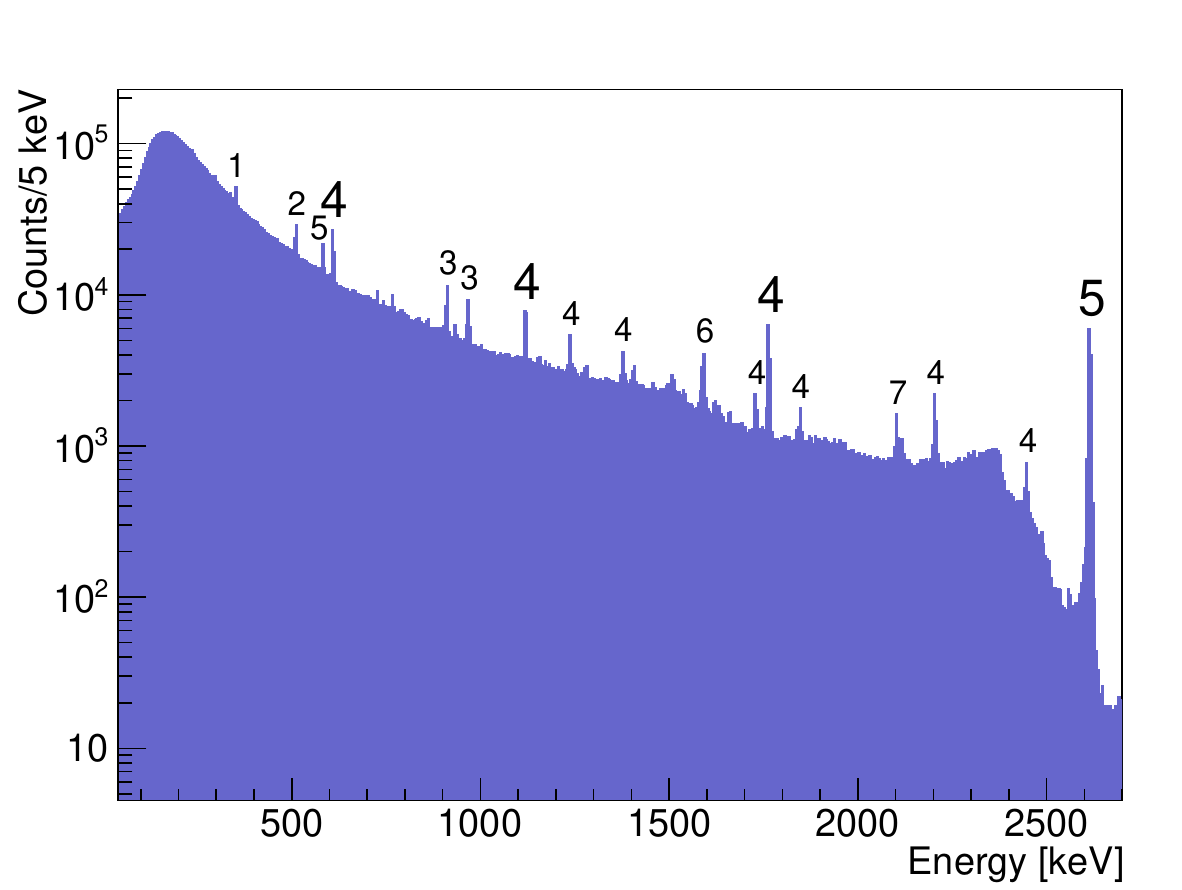}
    \includegraphics[width=0.49\textwidth]{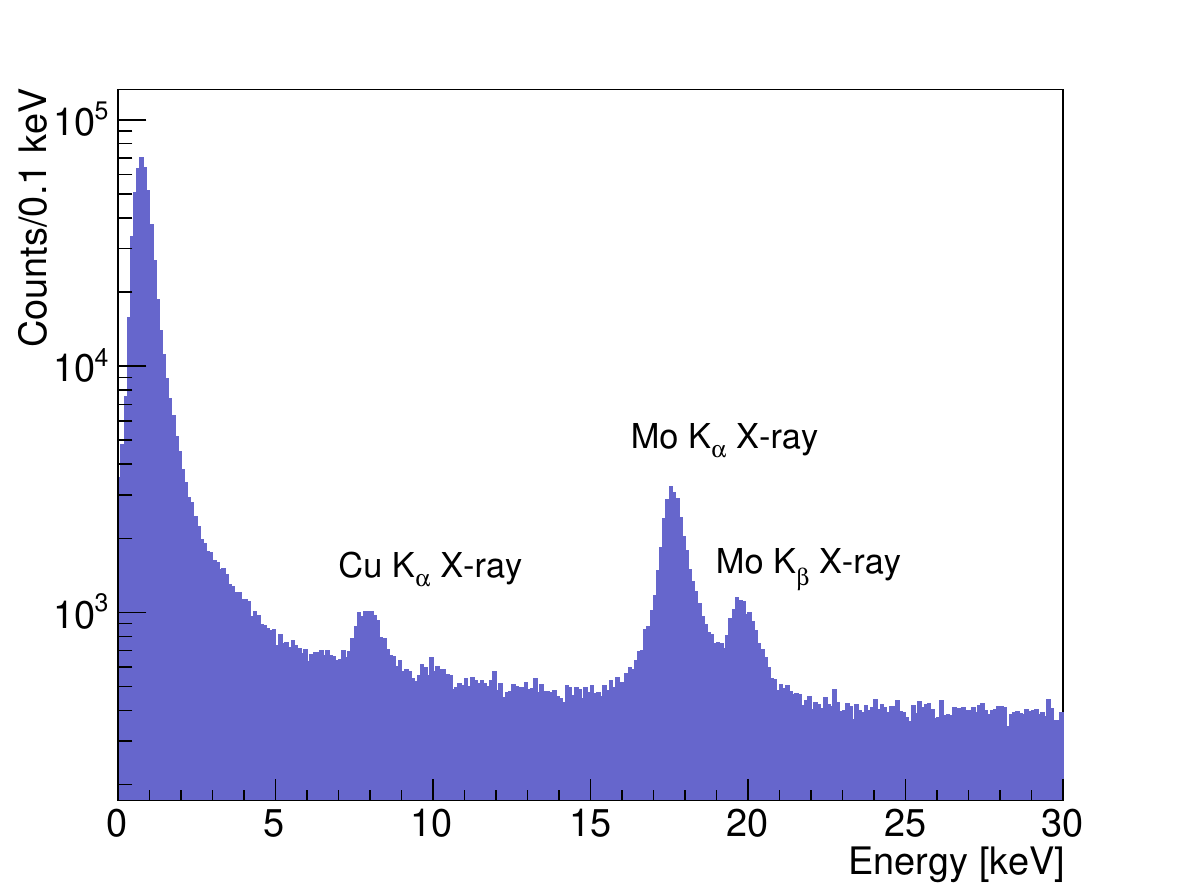}
    \caption{CUPID-Mo calibration spectra for both LMO detectors and LDs. \textit{Left:} Calibration spectra for LMO detectors exposed to the $^{232}$Th/$^{238}$U source. A selection of the most prominent peaks are labeled: (1) $^{214}$Pb, (2) $e^{-}e^{+}$ annihilation, (3) $^{228}$Ac, (4) $^{214}$Bi, (5) $^{208}$Tl, (6) the double escape peak for 2615~\,keV and (7) the single escape peak for 2615~\,keV $\gamma$'s. The four most prominent $\gamma$ peaks (denoted by larger labels) are utilized for LMO detector energy calibration. \textit{Right:} Calibration spectra for LDs with X-ray fluorescence during irradiation with a high activity $^{60}$Co source. The $\sim$17~\,keV X-ray line is used for a linear LD absolute energy calibration.}
    \label{fig:cal}
\end{figure*}
To perform energy calibration, four of the most prominent $\gamma$ peaks from the U/Th source are utilized: 609, 1120, 1764, and 2615~\,keV. These peaks are fit to a model comprised of a smeared-step function and linear component for the background, along with a crystal ball function~\cite{Oreglia1980} for the peak shape. The smeared step is modeled via a complimentary error function with mean and sigma equal to those used in the peak shape. Then, the best-fit peak location values are fit against the literature values for the specified $\gamma$ energies using a quadratic function with zero intercept which provides the calibration from the thermal gain corrected amplitude to energy for each channel:
\begin{equation}
    E(\tilde{C}) = p_{0}\tilde{C} + p_{1}\tilde{C}^{2}.
\end{equation}
In general this fit performs well for the selected $\gamma$ peaks used in calibration with only minimal residuals.
Using these calibration functions we can compute the deposited energy for each event, it is at this point that summed spectra from all channels can be meaningful for 0\vbb decay analysis. We note that between successive datasets, there is some small variation in the calibration fit function coefficients for any given channel, however this is acceptable as the calibration removes residual detector response  non-linearities that may change slightly over the course of the data taking. We check the stability of each calibration run over all datasets for each channel relative to the expected energy and find the central location of the 2615~\,keV $\gamma$ peak for each channel-run is consistent to within the channel energy resolution.
\subsection{LD Calibration}
\label{sec:ldcalib}
The LD energy scale is calibrated using a high activity $^{60}$Co source. This source produces 1173, 1333~\,keV $\gamma$'s which interact with the LMO crystals to produce fluorescence X-rays. In particular, Mo X-rays with energy $\sim$17~\,keV can be fully absorbed in the LDs and used for energy calibration. We use Monte Carlo simulations to determine the energy of the X-ray peak, accounting for the expected contribution of scintillation light. We extract the amplitude of the X-ray peak for each channel using a Gaussian fit with linear background and perform a linear calibration. Three datasets do not have any $^{60}$Co calibration available, so we assume a constant light yield with respect to the closest dataset in time that does have a $^{60}$Co calibration and extrapolate the LD calibration instead. The combined $^{60}$Co calibration spectrum is shown in Fig.~\ref{fig:cal}.

\subsection{Time Delay Correction}
\label{sec:jitter}
For studies that involve the use of timing information of events in multiple crystals, a correction of the characteristic time offsets between pairs of channels is performed. This correction is done by constructing a matrix of channel-channel time delays using $\gamma$ events that are coincident in two LMO detectors (referred to as multiplicity two, \mtwo) within a conservative ($\pm$100~\,ms) time window, and whose energy sum to a prominent $\gamma$ peak in the calibration spectra. This is done to ensure the events under consideration are likely to originate from causally related interactions and not from accidental coincidences.

\begin{figure}[htpb]
    \centering
    \includegraphics[width=\columnwidth]{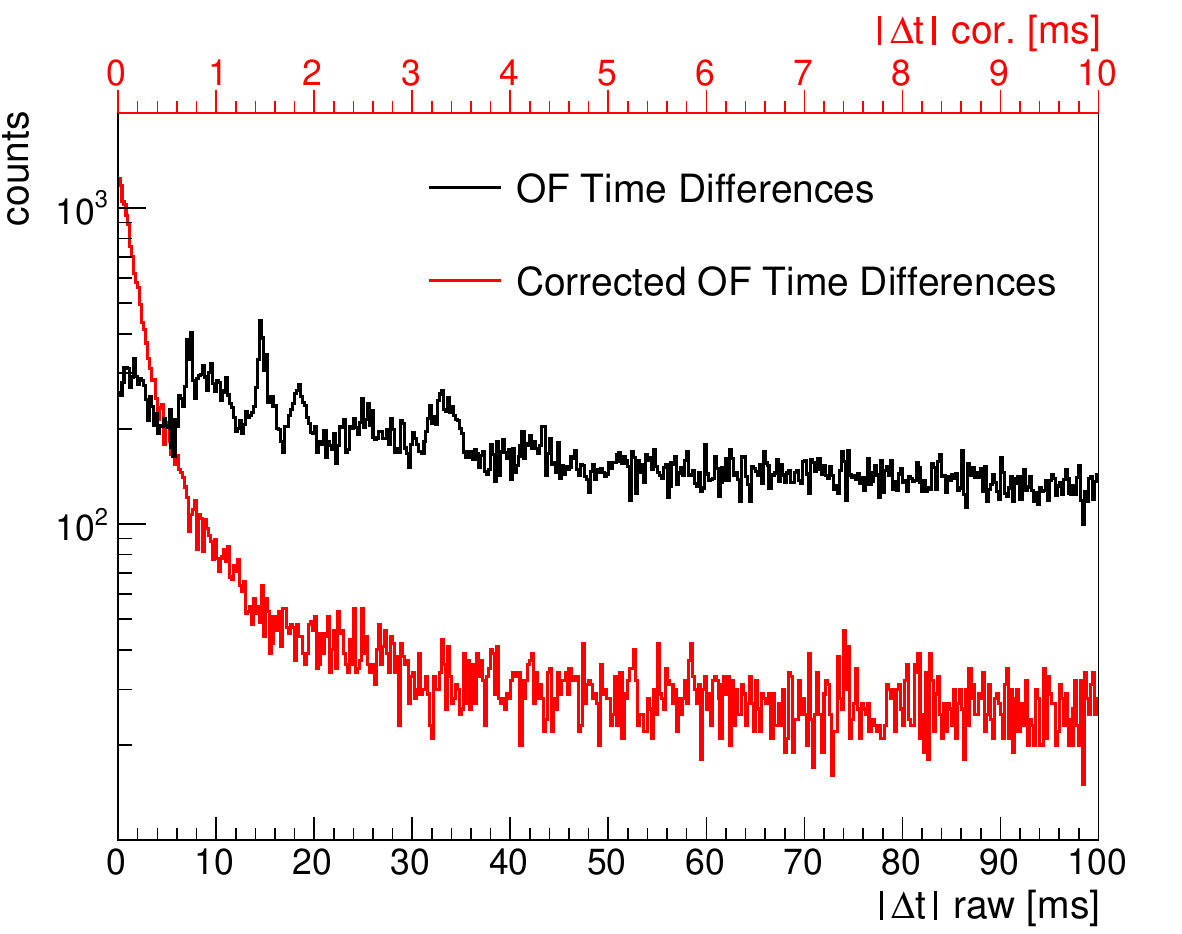}
    \caption{Time differences for $\mathcal{M}_2$ events whose energy sums to a prominent $\gamma$ peak in calibration data for both raw time (black) and corrected times (red). Note the time scales are different in the two cases to account for the much sharper peak with corrected times. Due to the high event rate in calibration data, an elevated rate of accidental coincidences is present leading to the presence of an elevated flat background in the $\Delta t$ distributions.}
    \label{fig:dt}
\end{figure}

The timing information for an event comes from two sources: the raw trigger time and an offset from the OF. The OF time, $t_{\text{OF},i}$, is the interpolated time offset which minimizes the $\chi^2$ between a pulse and the average pulse template. Together these two values are used to estimate the time differences between any two events, $i$ and $j$:
\begin{equation}
    \Delta t_{i,j} = (t_{\text{OF},i} + t_{\text{trig},i}) - (t_{\text{OF},j} + t_{\text{trig},j}).
    \label{eq:DeltaT}
\end{equation}
The distribution of this time offset for a given channel pair is computed. From this the time offset between channels $j,k$ ($\hat{t}_{j,k}$) is estimated as the median of the distribution. Several checks of the reliability of this estimate are performed: consistency of median and mode to within the $\sim$1\,ms binning size, and that there are sufficient counts ($\geq 5$). Any channel pair that fails either of these checks is deemed unsuitable for direct computation of $\hat{t}$ and an iterative approach is used exploiting the fact that time differences add linearly:
\begin{equation}
    \hat{t}_{i,j} = \hat{t}_{i,k} + \hat{t}_{k,j}.
\end{equation}
Several cross-checks for validity of the values in the time delay matrix are performed. \deltat values computed on the entire multiplicity two spectra are compared to those computed solely from the \mtwo summed $\gamma$ peaks and found to agree within $\sim$ 1\,ms. We purposefully zero out valid channel-pair cells in the matrix to check the reliability of the iterative approach, finding it reliably reproduces the \deltat values that are directly computable. As described in section~\ref{sec:coinc}, this time delay correction greatly improves our anti-coincidence cut as the distributions of corrected time differences is much narrower (see Fig.~\ref{fig:dt}).

\section{Data Selection Cuts and Blinding}
\label{sec:additionalProd}
After calibration is performed, the data are able to be meaningfully combined for analysis. We apply a set of simple ``base'' cuts to remove bad events. These cuts require that an event be flagged as a signal event (i.e, not a heater nor noise event), reject periods of bad detector operating conditions manually flagged due to excessive noise or environmental disturbances, reject any events with extremely atypical rise times, and reject any events with atypical baseline slope values. Additionally we reject all events from a single LMO that was observed to have an abnormally low signal-to-noise ratio which compromises its performance, as was done previously~\cite{CuMoPRL}. Beyond these base cuts, other improvements are possible with the use of more sophisticated selection cuts to remove background in order to increase the sensitivity to 0\vbb decay. We expect to observe background from:
\begin{itemize}
    \item spurious / pileup events, suppressed with pulse shape discrimination cuts (see section \ref{sec:psd});
    \item external $\gamma$ events, suppressed by removing multiple scatter events (see section \ref{sec:coinc});
    \item $\alpha $ background, removed using LD cuts (see section \ref{sec:light});
    \item $\beta$ events from close sources, suppressed by delayed coincidence cuts (see section \ref{sec:dc});
    \item external muon induced events, removed with muon veto (see section \ref{sec:muonveto}).
\end{itemize}
Finally, we note that all cuts are tuned without utilizing data in the vicinity of \qbb (3034~\,keV) for \isomo. As was done previously \cite{CuMoPRL}, we blind data by excluding all events in a 100~\,keV window centered at \qbb. In the following sections we describe these selection cuts.
\section{Pulse-Shape Discrimination}
\label{sec:psd}
An expected significant contribution to the background near \qbb are pileup events in which two or more events overlap in time in the same LMO detector. This causes incorrect amplitude estimation and shifts events into our region of interest (ROI). In order to mitigate this effect we employ a pulse shape discrimination (PSD) cut that is comprised of two different techniques.

The main method we utilize for pulse-shape discrimination is based on principal component analysis (PCA), as was originally utilized in the previous analysis~\cite{CuMoPRL, PCA_Huang_2021}, and successfully applied recently to \Q~\cite{CUORE1ton} with more details in~\cite{Huang2021}. This method utilizes 2\vbb decay events between 1--2~\,MeV to derive a set of principal components that are used to describe typical pulse shapes for each channel-dataset. The leading principal component typically resembles an average pulse template with subsequent components adding small adjustments. These are used to compute a quantity referred to as the reconstruction error ($RE$) which characterizes how well a given pulse with $n$ samples, $\bm{x}$, is described by a set of $m$ principal components:
\begin{equation}
    RE = \sqrt{\sum_{i=1}^{n}{\left(x_{i} - \sum_{k=1}^{m}{q_{k}w_{k,i}}\right)^{2}}},
\end{equation}
where $\bm{w_{k}}$ is the $k$-th eigenvector of the PCA with the projection of $\bm{x}$ onto each component given by $q_{k} = \bm{x}\cdot\bm{w}_{k}$. $RE$ is energy dependent and this is corrected for by subtracting the linear component, $f(E)$, and normalizing by the median absolute deviation (MAD):
\begin{equation}
    NE = \frac{RE - f(E)}{MAD}.
\end{equation}
The resulting normalized reconstruction error, $NE$, is then used with an energy independent threshold to reject abnormal events.

\subsection{PCA Improvements}
\label{sec:pca}
We improve several aspects of the PCA cut compared to the previous implementation \cite{PCA_Huang_2021}: we utilize a cleaner training sample, perform normalization on a run-by-run basis, and correct for the energy dependence of the MAD. Abnormal pulses in the training sample result in distortions to all principal components leading to degraded performance in both efficiency and rejection power. To mitigate this we use a stricter selection cut requiring that the pretrigger baseline RMS not be identically zero (indicative of digitizer saturation and subsequent baseline jumps), and that a simple pulse counting algorithm must identify no more than one pulse on the LMO waveform and primary LD in the event window. This cleaner training sample allows us to utilize higher numbers of principal components without sacrificing efficiency.
\begin{figure*}[ht!]
    \centering
    \begin{minipage}{0.6\textwidth}
        \includegraphics[width=\linewidth]{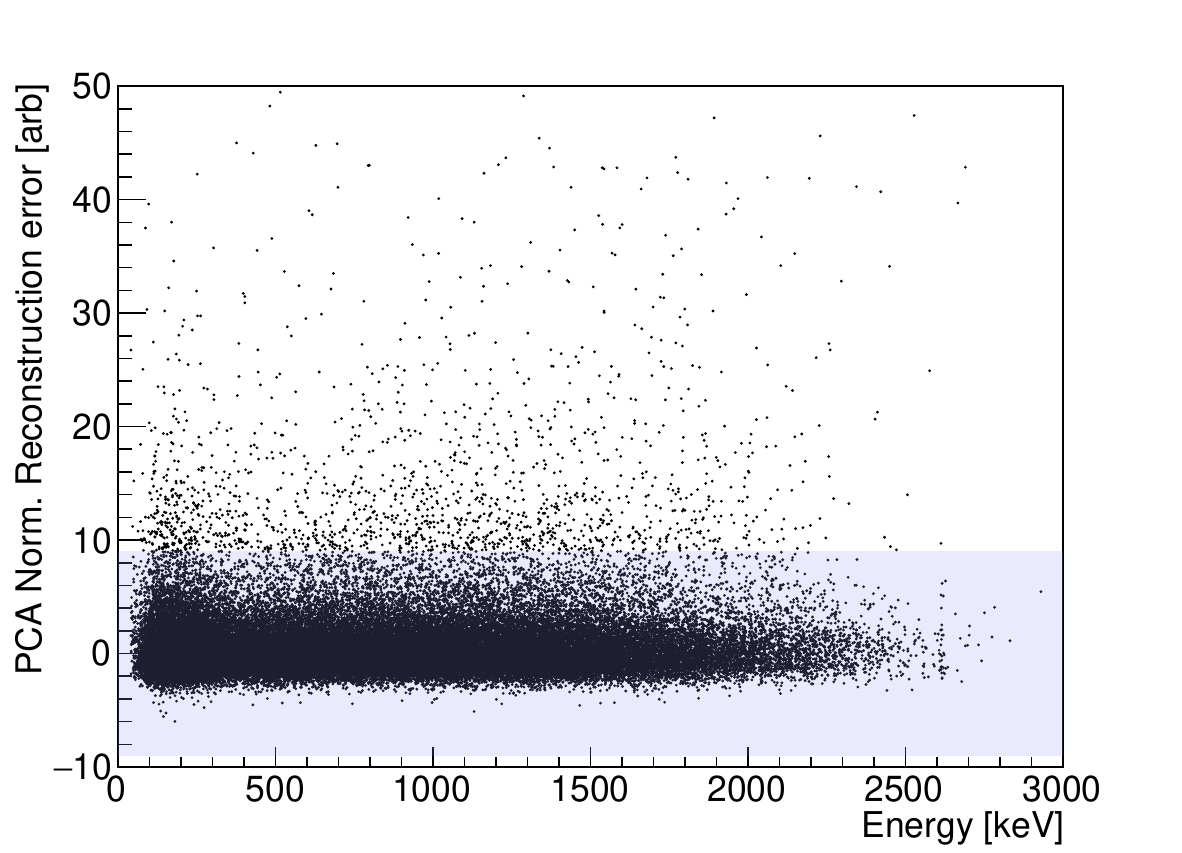}
    \end{minipage}
    \begin{minipage}{0.38\textwidth}
        \includegraphics[width=\linewidth]{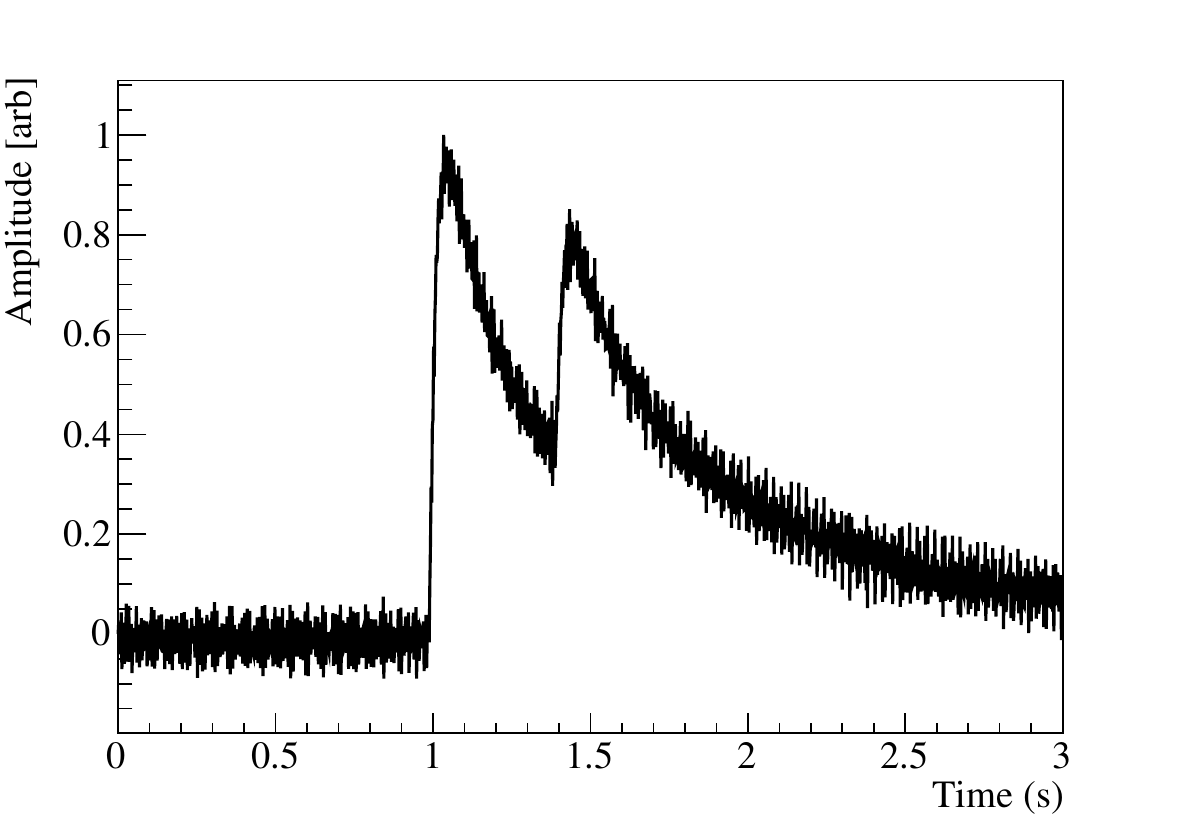}
        \includegraphics[width=\linewidth]{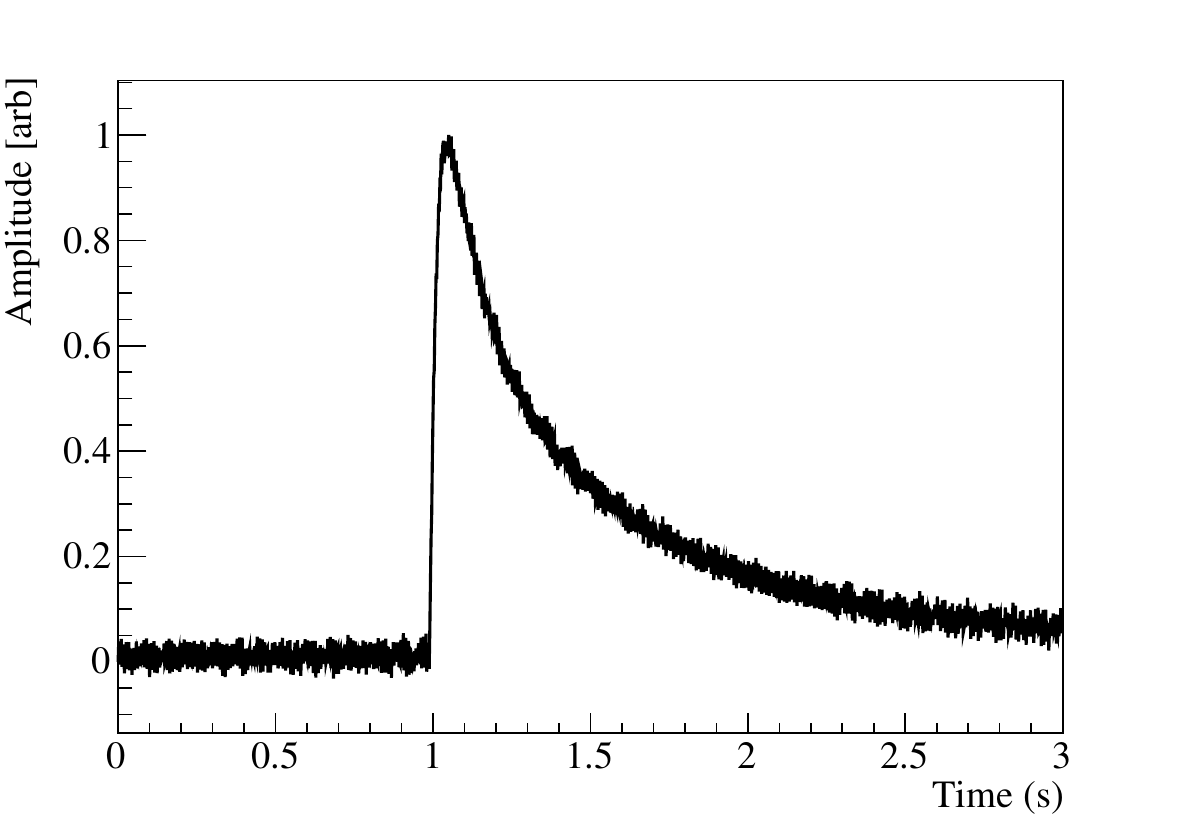}
    \end{minipage}
    \caption{Example of the normalized PCA reconstruction error (\textit{left}) as a function of energy, and two types of events (\textit{right}) that exist in our data. (\textit{left:}) The normalized reconstruction errors have no energy dependence and are normalized on a run-by-run basis for each channel allowing for a single energy independent cut to be applied across all channels in every dataset. Events with higher normalized reconstruction errors are removed and likely have incorrect energy values which may cause such events to be shifted into the region of interest. The shaded box denotes the acceptable normalized reconstruction error range for this cut. (\textit{right}): The top event is an example of a pileup event with high normalized reconstruction error ($\sim$22) at $\sim$1189~\,keV which would have an incorrect amplitude reconstruction. The bottom event is a more typical pulse with a small normalized reconstruction error ($\sim$0.5) at $\sim$2482~\,keV, having no resulting error in its OF amplitude reconstruction.
    }
     \label{fig:pcaFail}
\end{figure*}
By performing the normalization of $RE$ on a run-by-run basis, as opposed to whole-dataset, the fit for the linear component better reflects changes in $RE$ that may arise due to variations in noise. To correct for the energy dependence of the MAD, we require the aggregate statistics of a whole dataset. We perform a linear regression in energy and compute the average MAD of the ensemble. We then use the ratio of the linear regression function and ensemble average MAD as a correction to the individual channel MAD values, providing a proxy for a channel-dependent energy scaling of the MAD.

We examine the overall efficiency, impact on the 2615~\,keV $\gamma$ peak resolution, and optimization of the median discovery significance as suggested in Cowan et al.~\cite{CowanMetric}, as a function of number of PCA components and cut threshold. From this we choose to utilize the first 6 leading components of the PCA for this portion of the PSD cut. As seen in Fig.~\ref{fig:pcaFail} the $NE$ quantity has no energy dependence and is able to reject obvious abnormal pulses.

\subsection{PSD Enhancements}
\label{sec:psd_enhance}
To finalize the PSD cut we utilize a two additional parameters developed in previous \Q analyses~\cite{CUOREPRL2017}. These parameters are computed on the optimally filtered pulse itself and are measures of goodness of fit on the left/right side of the filtered pulse, and are referred to as test-value-left and test-value-right (TVL and TVR) respectively. These $\chi^2$-like quantities are normalized via empirical fits of their median and MAD energy dependencies using $\gamma$ events between 500--2600~\,keV. As these quantities are computed on the filtered pulses they provide an additional proxy to detect subtle pulse-shape deviations and provide a complimentary way to reject pileup events, especially for noisy events~\cite{PhysRevC.104.015501}.

We observe that some pileup events still leak through the six component PCA cut alone, primarily pileup with a short separation with the earlier pulse having a small amplitude relative to the ``primary'' pulse. Energy independent cuts on TVL and TVR are able to remove a large portion of these with negligible loss to efficiency. The discrimination power from these two cuts arises from the fact they are derived on the optimally filtered waveforms. They are sensitive to pileup in a fashion that the PCA is not, and owing to the better signal-to-noise ratio, tend reject small-scale pileup events that the PCA cut is insensitive to. We combine the various pulse-shape cuts to form the final PSD cut by requiring that the absolute value of the normalized reconstruction error be less than 9, and that the absolute value of the normalized TVR and TVL quantities each be less than 10. The resulting cut maintains an efficiency comparable to the previous analysis (see section~\ref{sec:eff}) while being able to reject more types of abnormal events.
\section{Anti-Coincidence}
\label{sec:coinc}
Due to the short range of \isomo $\beta\beta$ electrons in LMO (up to a few\,mm~\cite{BandacBetaDepth}), 0\vbb decay events would primarily be contained within in a single crystal. A powerful tool to reduce backgrounds is to remove events where simultaneous energy deposits in multiple LMO crystals occur. It is useful to classify multi-crystal events for a background model and other analyses (e.g. $\beta\beta$ transitions to excited states). We define the multiplicity, $\mathcal{M}_i$, of an event by the total number of coincident crystals with an energy above 40~\,keV in a pre-determined time window. This requires measuring the relative times of events across different crystals. Previously we utilized a very conservative window of $\pm$100~\,ms, which due to the relatively fast 2\vbb decay rate in \isomo of $\sim7\times10^{18}$~\,yr or $\sim$2~\,mHz in a 0.2~\,kg $^{100}$Mo-enriched LMO crystal~\cite{CUPIDMo.2nu.precise}, leads to $\sim$2\% of single crystal (\mone) events being accidentally tagged as two-crystal (\mtwo) events. This results in a slight pollution of the \mtwo energy spectrum with these random coincidences as events that should be \mone have been incorrectly tagged as \mtwo events. The channel-channel time offset correction described in section~\ref{sec:jitter} substantially narrows the $\Delta t$ distribution amongst channel-channel pairs allowing for a much shorter time window to be used (see Fig.~\ref{fig:dt}). For this analysis we choose a coincidence window of 10\,ms which reduces the dead time due to accidental tagging of \mone events as \mtwo by a factor of $\sim$10, while also producing a more pure \mtwo spectrum. The anti-coincidence (AC) cut then ensures we only examine single-crystal events. 
\section{Light Yield}
\label{sec:light}

\begin{figure*}[htpb]
    \centering
    \includegraphics[width=0.49\textwidth]{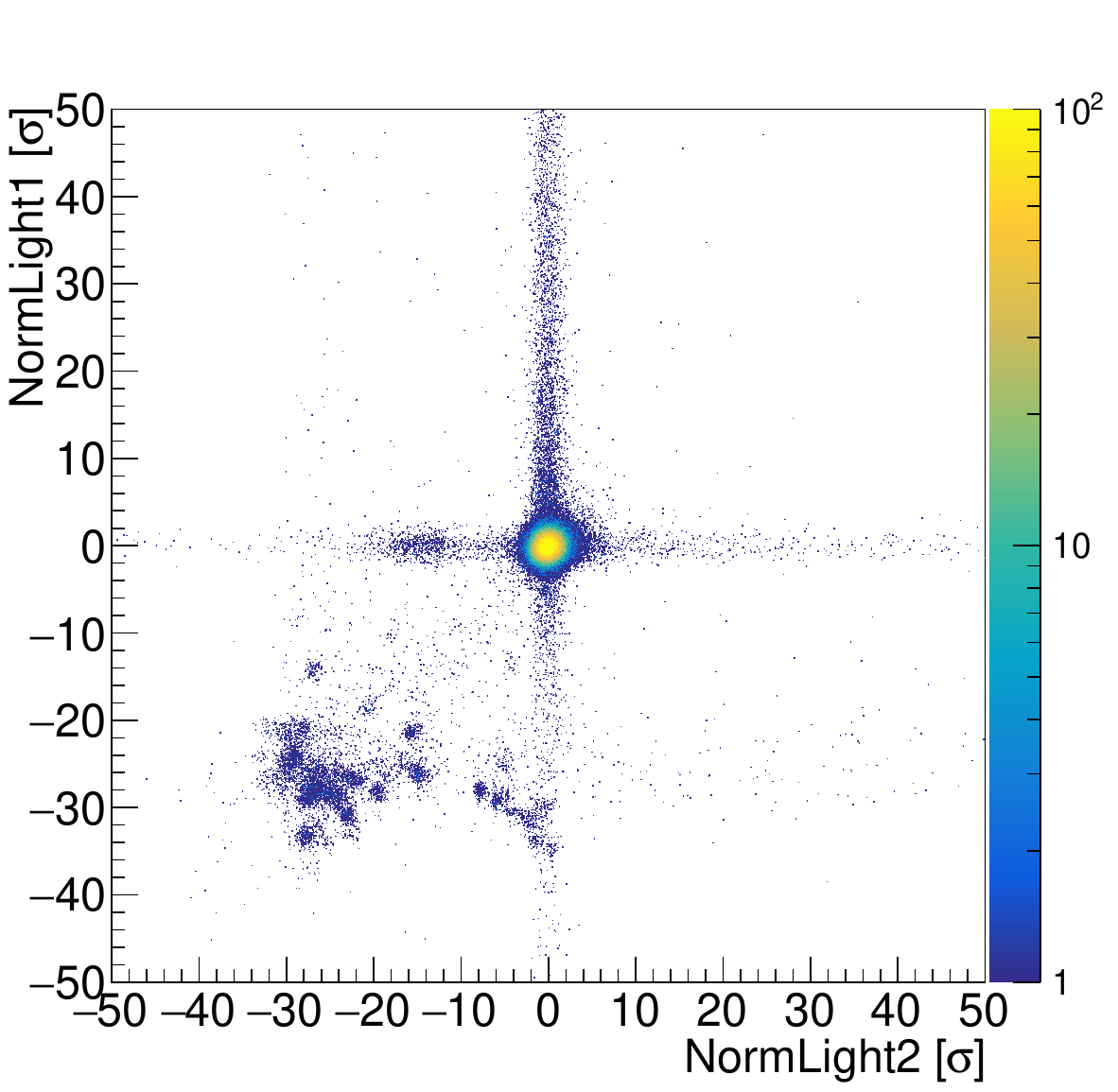}
    \includegraphics[width=0.49\textwidth]{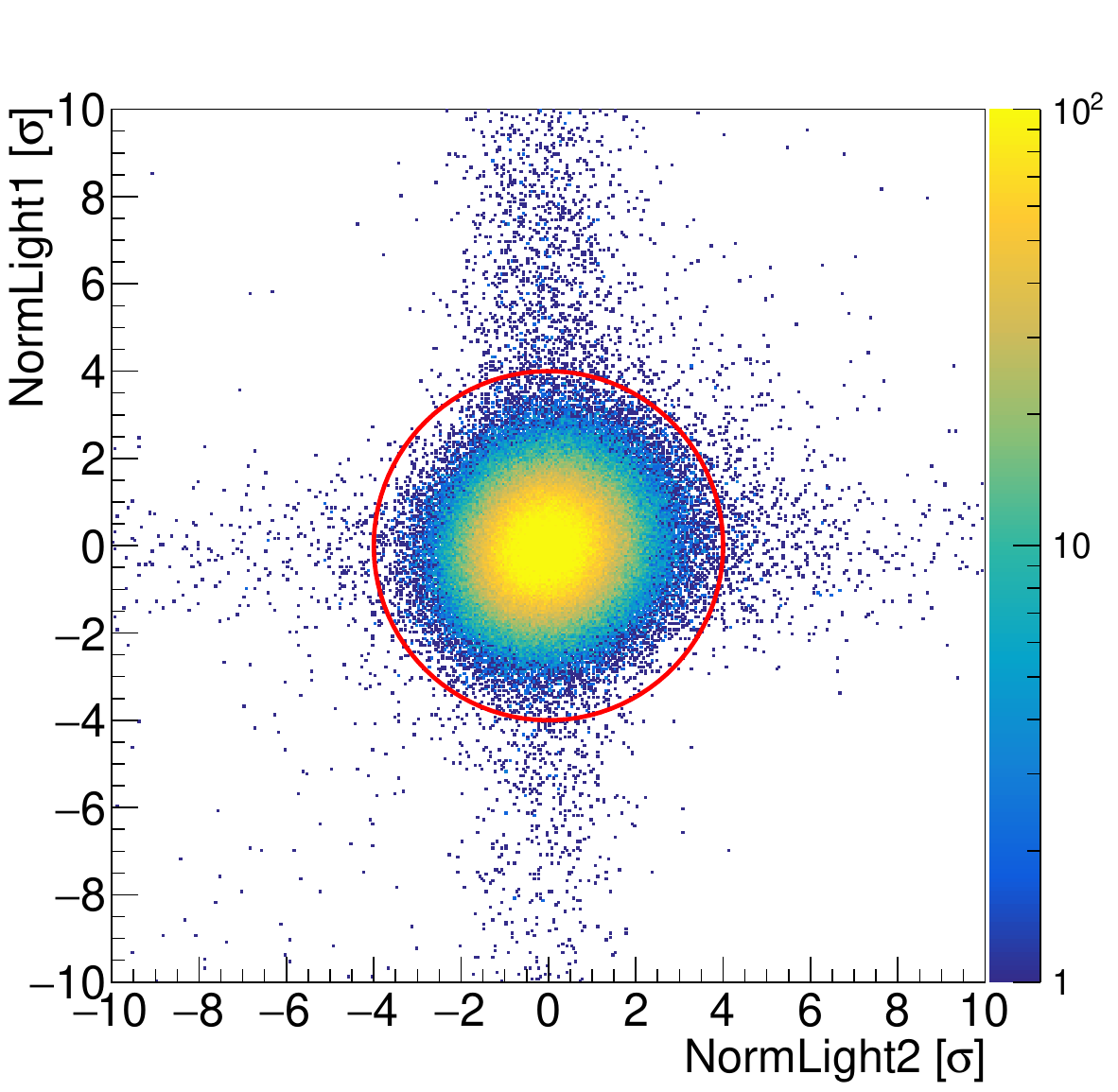}
    \includegraphics[width=0.99\textwidth]{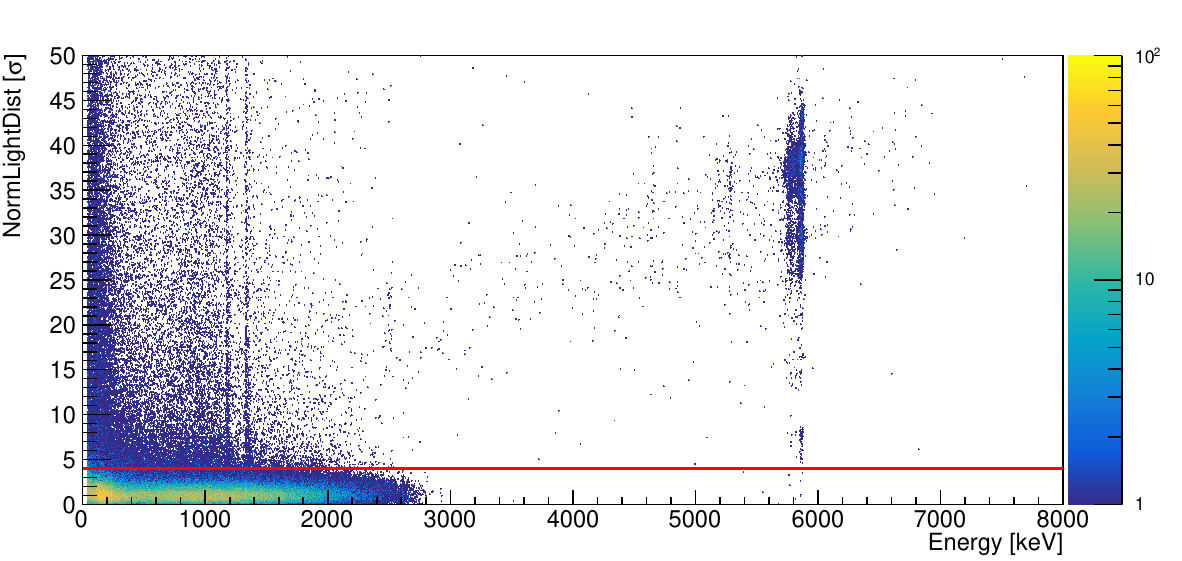}
    \caption{Two dimensional distribution of normalized light variables with LMO detector energy >1~\,MeV, zoomed out ({\it top, left}) and zoomed in ({\it top, right}) with the LY cut definition of $D<4$ also shown (solid circle). We observe that $\gamma/\beta$ signal like events are distributed around (0,0). The vertical band in the left figure is the result of a single light detector that has an excess $^{60}$Co contamination resulting in a higher rate of events that deposit significant energy into this light detector. These are easily rejected with the light cut shown here as well as an anti-coincidence cut with the specific LD. Events populating the lower left quadrant are $\alpha$'s from the various detectors and show the hallmark deficiency of scintillation light on both light sensors. The energy dependence of the normalized light distance is shown in the bottom figure, again with the cut boundary denoted by a solid line. $\alpha$ events are clearly well separated from the $\beta$/$\gamma$'s. We observe that the norm light distance for these events is roughly flat with energy. The contamination due to excess $^{60}$Co on 1 LD is evident. A cluster of $\alpha$ events at low normalized light distance is present due to a short period of time with sub-optimal performance on a single LD.}
    \label{fig:NL_plots}
\end{figure*}

LDs are the primary tool we use in \cumo to distinguish $\alpha$ from \betagamma particles to reduce degraded $\alpha$ backgrounds. Using the detected LD signal relative to energy deposited in the LMO detector, we are able to separate $\alpha$'s from \betagamma events as the former have $\sim$20\% the light yield of the latter for the same heat energy release. Previously, we exploited the information provided from the LDs by using a resolution-weighted summed quantity and direct difference to select events with light signals consistent with \betagamma's~\cite{CuMoPRL}. In this analysis we modify the light cuts to utilize the correlation between both LDs associated with an LMO detector more directly.
To account for the energy dependence of the light cut, we model the light band mean and width. We divide the light band into slices in energy for each channel and dataset. For each slice we perform a Gaussian fit of the LD energies to determine the mean and resolution, then fit the means to a second order polynomial in energy, and the resolutions to: 
\begin{equation}
    \sigma(E) = \sqrt{p_0^2+p_1\cdot E}.
\end{equation}
This is used to determine the best estimate of the expected LD energy for a given energy. We define the normalized LD energy for a given LMO detector $c$ in dataset $d$ as:
\begin{equation}
    n_{c,s,d} = \frac{L_{c,s,d}-\widehat{L}_{c,s,d}(E)}{\sigma(E)_{c,d,s}},
\end{equation}
where $s$ is the LD neighbor index, $L_{c,s,d}$ is the measured LD energy, $\widehat{L}_{c,s,d}(E)$ is the expected LD energy, and $\sigma_{c,d,s}(E)$ is the expected width of the light band. This procedure explicitly removes the energy dependence, and we note that $n_{c,s,d}$ has a normal distribution.

We expect signal-like \betagamma events to have similar energies on the both LDs~\cite{CuMo_instrument}. We observe background events where the total light energy is consistent with \betagamma signal events but the resulting individual LD energies are very different. This can happen due to surface $\alpha$ events where a nuclear recoil deposits some energy onto only one LD (see~\cite{lumineu2017}), or contamination on the LDs themselves. To remove these background-like events we exploit the full information of two LDs by making a two-dimensional light cut. In particular, we expect the joint distribution of $n_{c,0,d}$ and $n_{c,1,d}$ to be a bivariate Gaussian. This is also observed in data, with minimal correlations between the two normalized LD energies, thus a simple radial cut can be defined by computing the normalized light distance, $D_{c,d}$:
\begin{equation}
    D_{c,d} = \sqrt{n_{c,0,d}^2+n_{c,1,d}^2}.
\end{equation}
For channels which do not have two LDs we instead make a simple cut on the single normalised light energy which is available. We chose a cut of $D<4$ (corresponding to $\sim 3.5  \sigma$ equivalent coverage). As shown in Fig.~\ref{fig:NL_plots} this is sufficient to remove the $\alpha$ background which is characterized by a large negative value of $n_{c,s,d}$. 
\section{Delayed Coincidences}
\label{sec:dc}
A significant background for calorimeters can be surface and bulk activity in the crystals themselves due to natural U/Th radioactivity (see~\cite{Denys.Andrea.LTD} for more details). In particular, because $Q_{\beta\beta}$ of \isomo (3034~\,keV) is above most natural radioactivity, the only potentially relevant isotopes are $^{208}$Tl, $^{210}$Tl and $^{214}$Bi~\cite{PirroScintBolo}. However, given both the low contamination in the \cumo detectors and the very small branching ratio ($\sim$0.02\%), the decay chain of $^{214}$Bi $\rightarrow$ $^{210}$Tl $\rightarrow$ $^{210}$Pb is negligible.

For $^{208}$Tl the decay chain proceeds as:
\begin{equation}
    ^{212}\text{Bi}\xLongrightarrow[\alpha (6207~\,\text{keV})]{60.6~\,\text{min}, \ \ 35.9\% \ \text{BR}}\ ^{208}\text{Tl}\xLongrightarrow[\beta^{-} (4999~\,\text{keV})]{3.1~\,\text{min}, \ \ 100\% \ \text{BR}}\ ^{208}\text{Pb}.
\end{equation}
A common approach is to reject candidate $^{208}$Tl events that are preceded by a $^{212}$Bi $\alpha$ decay~\cite{PirroScintBolo,CUPID0Final}. We note that for bulk activity, the candidate $\alpha$ is detected with $>99$\% probability, so it is the efficiency at which these $\alpha$ events pass the analysis cuts that sets this background. For surface $\alpha$ events, $\sim$50\% reconstruct at their $Q$-value, so a delayed coincidence cut would remove only about $\sim$50\% of surface events (see \cite{CUPID0Final}). In this analysis we use the same energy and time difference as was used previously~\cite{CuMoPRL}: we reject any candidate $^{208}$Tl event that is within 10 half-lives from a $^{212}$Bi candidate event. We note that the \cumo detector structure with a reflective foil and Cu holder surrounding each crystal reduces the effectiveness of this cut for surface events. In a future experiment with an open structure (for example CUPID~\cite{CUPID:2019}, CROSS~\cite{CROSS}, or BINGO~\cite{Nones2021}) the detection of multi-site $\alpha$ events may significantly improve this detection probability (and therefore cut rejection). 
\begin{table}[htpb]
    \centering
    \caption{Energy and time selections used for CUPID-Mo delayed coincidence cuts}
    \begin{tabular}{llll}
    Cut     & $Q_{\alpha} \ \mathrm{[keV]}$ & Energy cut [keV] & Time [s] \\ \hline \hline
    $^{212}$Bi\ $^{208}$Tl   & $6207$    &  $6000-6300$ & 1830 \\
    $^{222}$Rn\ $^{214}$Bi &  $5590$ & $5490-5620$    & 13860       \\
    $^{218}$Po\ $^{214}$Bi & $6115$ & $6015-6565$ & 13620
    \end{tabular}
    \label{tab:DC}
\end{table}

In addition to this commonly used cut, the extremely low count rate for $\alpha$'s in \cumo, due to low contamination~\cite{Schmidt2020, Poda2020}, enables a novel extended delayed-coincidence cut designed to remove potential $^{214}$Bi induced events. We focus on the lower part of the decay chain:
\begin{equation}
\begin{split}
    &^{222}\text{Rn}\xLongrightarrow[\alpha (5590~\,\text{keV})]{3.8~\,\text{day}, \ \ 100\% \ \text{BR}}\ ^{218}\text{Po}\xLongrightarrow[\alpha (6115~\,\text{keV})]{3.1~\,\text{min}, \ \ 99.98\% \ \text{BR}}\ ^{214}\text{Pb} \\ &^{214}\text{Pb}\xLongrightarrow[\beta^{-} (1018~\,\text{keV})]{27.1~\,\text{min}, \ \ 100\% \ \text{BR}}\ ^{214}\text{Bi}\xLongrightarrow[\beta^{-} (3269~\,\text{keV})]{19.7~\,\text{min}, \ \ 99.98\% \ \text{BR}}\ ^{214}\text{Po} \\ &^{214}\text{Po}\xLongrightarrow[\alpha (7834~\,\text{keV})]{163.4~\,\mu\text{s}, \ \ 100\% \ \text{BR}}\ ^{210}\text{Pb}
\end{split}
\end{equation}
We tag the $^{214}$Bi nuclei based on either the $^{222}$Rn or $^{218}$Po $\alpha$ decay. Compared to $^{212}$Bi $\rightarrow$ $^{208}$Tl coincidences, a much larger veto time window is required. We set these time cuts based on a simulation of the time differences between decays in order to have a 99\% probability of the decay being in the selected time range, as shown in Table~\ref{tab:DC}. We veto events where there is an $\alpha$ candidate within [$Q_{\alpha}-$100, $Q_{\alpha}$+50]~\,keV and within the time differences in Table \ref{tab:DC} in the same LMO detector. This energy range is chosen to fully cover the $Q$-value peaks. Despite the dead time per event being large, the total dead time is acceptable ($<1\%$, see section~\ref{sec:eff}) thanks to the low contamination of $^{226}$Ra in the \cumo detectors. We observe several events with $E>2600$ keV that are rejected, while the events removed at lower energy are dominated by accidental coincidences of 2\vbb decays. 
\section{Muon veto coincidences}
\label{sec:muonveto}
\begin{figure}[htpb]
    \centering
    \includegraphics[width=\columnwidth]{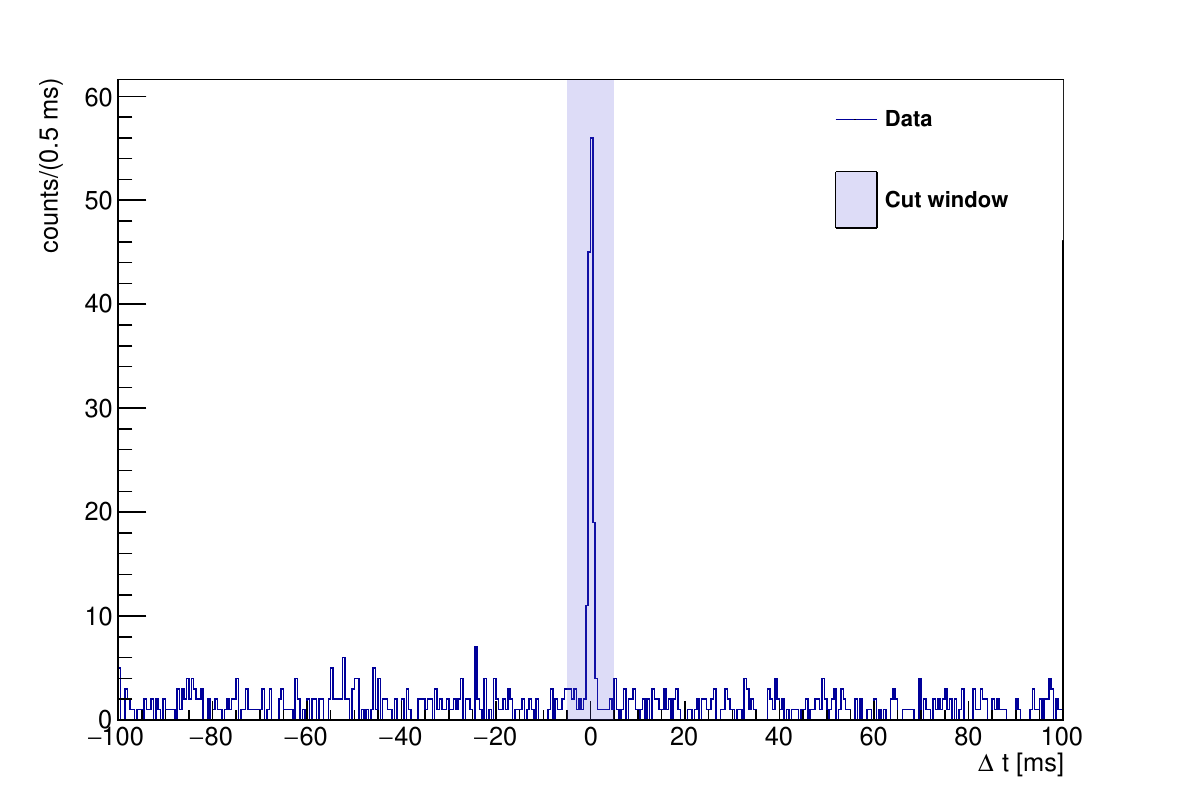}
    \caption{Time differences between muon veto and LMO detector events after we have subtracted off the peak offset ($\sim 60$ ms). The cut of $\pm 5$ ms is used in our analysis. To obtain a good signal-to-noise ratio, only events with $\mathcal{M}>1$ in the LMO detectors are selected.}
    \label{fig:muonveto}
\end{figure}
We apply an anti-coincidence cut between the LMO detectors and an active muon veto to reject prompt backgrounds from cosmic-ray muons which may deposit energy in the ROI, with LY similar to a $\gamma/\beta$. The muon veto system is described in detail in \cite{SCHMIDT201328}. We utilize muon veto timestamps to compute an initial set of coincidences between LMO detectors and the veto system. We observe a clear $\Delta t$ peak of muon induced events which we correct for (see Fig.~\ref{fig:muonveto}). The muon veto coincidences are then defined using the corrected times with a window of $\pm$ 5~\,ms. The relatively small window removes the need to also place a requirement on the number of muon veto panels triggered, maximizing the rejection of background events with minimal impact on livetime.

\section{Energy spectra}
\label{sec:other}

\begin{figure*}[htpb!]
    \centering
    \includegraphics[width=0.99\textwidth]{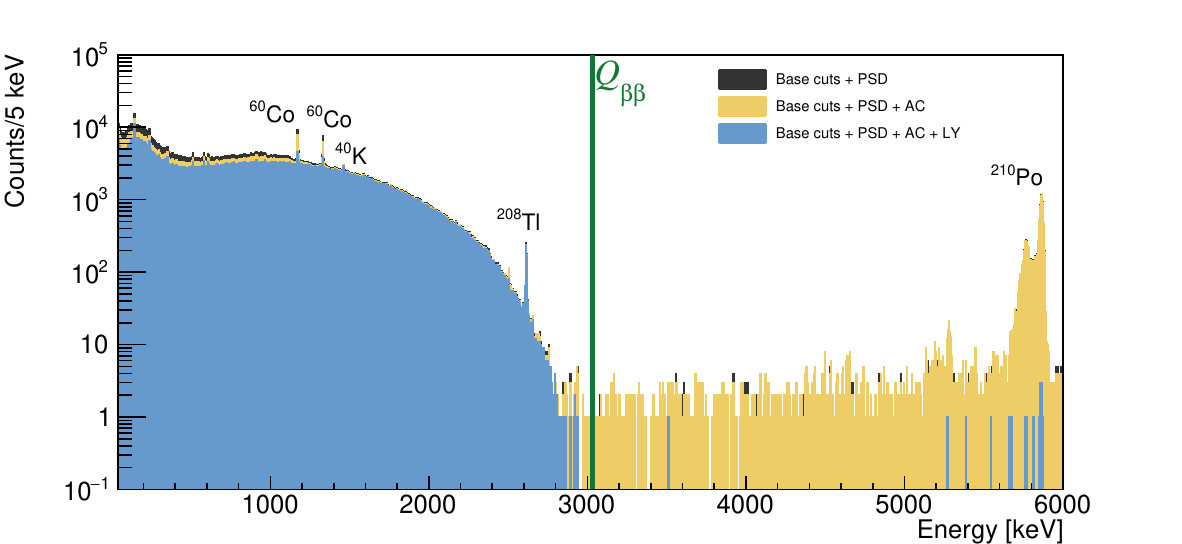}
    \caption{Unblinded spectrum of physics data in CUPID-Mo with successive application of cuts. The application of the PSD cuts produces a spectra containing all events with good pulse shape characteristics. Application of the anti-coincident (AC) cut removes all events with multiplicity greater than one, along with any possible delayed coincidences from $^{212}$Bi, $^{222}$Rn, or $^{218}$Po decay chains, and any events coincident with the muon veto. The final application of the light cut removes almost all the high energy background induced from degraded $\alpha$'s that remains, leaving a few intermediate LY events at high energy. The large green vertical line indicates the location of \qbb (3034~\,keV).}
    \label{fig:cutflow}
\end{figure*}

After all cuts are tuned on the blinded data we proceed to compute cut efficiencies, extract the resolution energy scaling, energy bias, and define the ROI. The application of successive cuts can be seen in Fig.~\ref{fig:cutflow}. Starting with the base cuts, the application of the PSD cuts produces a spectrum of events originating from real physical interactions with the detector (i.e., devoid of abnormal events). We see that the spectrum is dominated by 2\vbb decay from $\sim$1\,MeV up towards \qbb with few events populating the $\alpha$ region. The application of the AC cuts removes only a small amount of events as the majority of events are single-crystal interactions. The most significant selection cut is the application of the LY cut which removes almost all remaining events at high energies where degraded $\alpha$ events may be present.

\section{Efficiencies}
\label{sec:eff}
In order to compute the cut efficiencies we use three methods that span the distinct types of cuts present in this analysis:
\begin{itemize}
    \item noise events for pileup efficiency;
    \item efficiency from $\gamma$ peaks;
    \item efficiency from $^{210}$Po peak.
\end{itemize}
\begin{table}[htbp]
    \centering
    \caption{Efficiencies for \cumo selection cuts, evaluated either as a constant efficiency or linearly extrapolated to $Q_{\beta\beta}$. Methods used to compute each efficiency are indicated (see text)}
    \begin{tabular}{lcc}
        Cut &  Efficiency Flat / $Q_{\beta\beta}$ [\%] & Method \\ \hline \hline 
        Pileup & $95.7\pm 1.0$ & Noise \\
        Anti-coincidence & $99.55\pm0.07$ & $^{210}$Po \\
        Muon veto & $99.62\pm 0.07$ & $^{210}$Po\\
        Delayed coincidence & $99.16\pm 0.01$ & $^{210}$Po\\
        PSD & $95.2\pm 0.5/94.3\pm 1.5$ & $\gamma$ peaks\\
        Light Distance& $99.4\pm 0.4/99.7\pm 0.8$ & $\gamma$ peaks \\
        \hline
        Total Analysis Efficiency & \analysiseff & - \\
    \end{tabular}
    \label{tab:effs}
\end{table}
We note that the trigger efficiency for this analysis is taken as 100\%. The typical 90\% trigger thresholds are $\sim$8.5~\,keV and $\sim$0.55~\,keV for LMO detectors and LD's respectively, well below the 40~\,keV analysis threshold used by the anti-coincidence cuts. The trigger efficiencies are measured by injecting scaled pulse templates into actual noise events and running these through the optimum trigger for each channel-dataset pair. More details of this process are described in \cite{helisThesis} (Sect. 3.3.2).

The pileup efficiency is the probability that an event will not have another pulse in the same time window during which event reconstruction takes place. In addition, we check if the energy of the noise event is biased by $>$ 20 keV. If either of these two possibilities occur, we consider the event a pileup. We compute the pileup rejection efficiency as the ratio of the noise events passing the single trigger criterion and with energy inside $\pm$20~\,keV to the total number of noise events. We present the exposure weighted average over all datasets in Table~\ref{tab:effs} and assign a 1\% uncertainty to this calculation due to the extrapolation from noise to physics events. We note that this is equivalent to a statistical calculation based on the known trigger rate, but this method averages over varying trigger rates (in time or across channels).

The anti-coincidence, delayed coincidence, and muon veto cuts are not expected to have energy dependent efficiencies and represent detector deadtimes. 
For each of these we evaluate the efficiency utilizing events in the $^{210}$Po $Q$-value peak at 5407~\,keV, as this peak has a very high energy and provides a clean sample of physical events. We extract the efficiency as $\varepsilon=N_{\text{pass}}/N_{\text{total}}$ integrating in a $\pm$50\,keV window around the peak; the results are listed in Table~\ref{tab:effs}.

\begin{figure}[htpb!]
    \centering
    \includegraphics[width=\columnwidth]{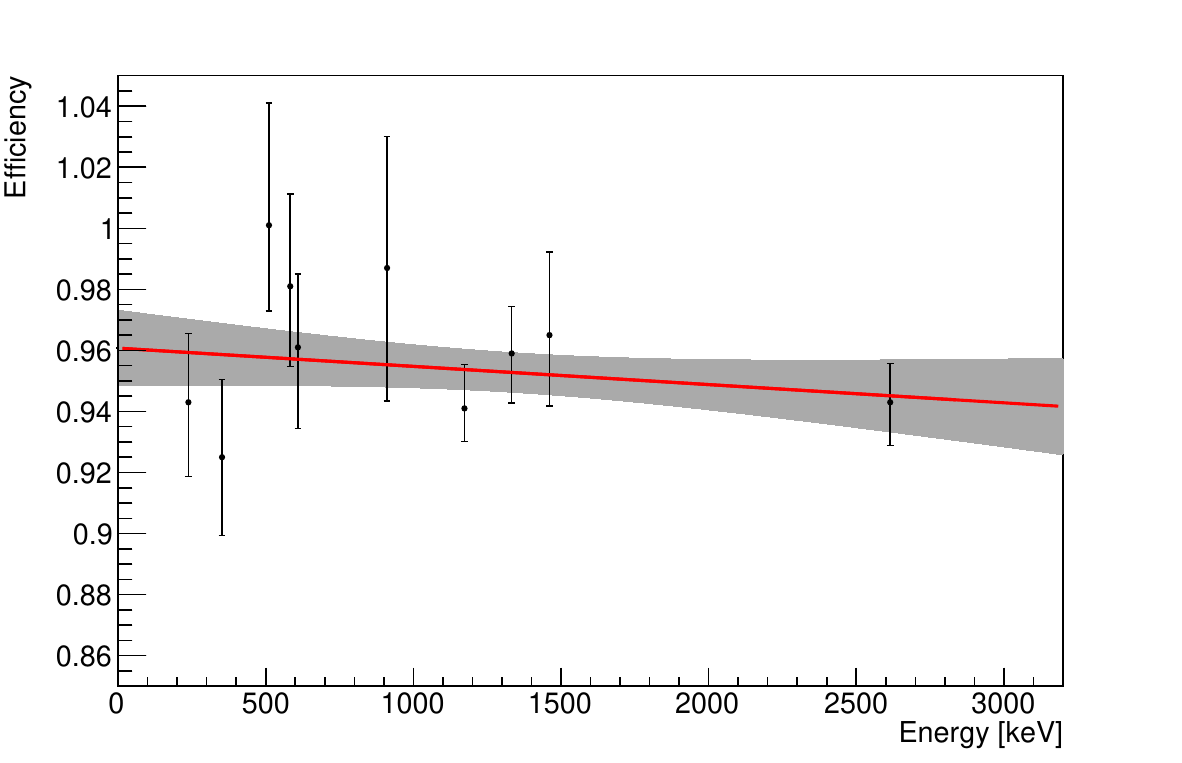}
    \includegraphics[width=\columnwidth]{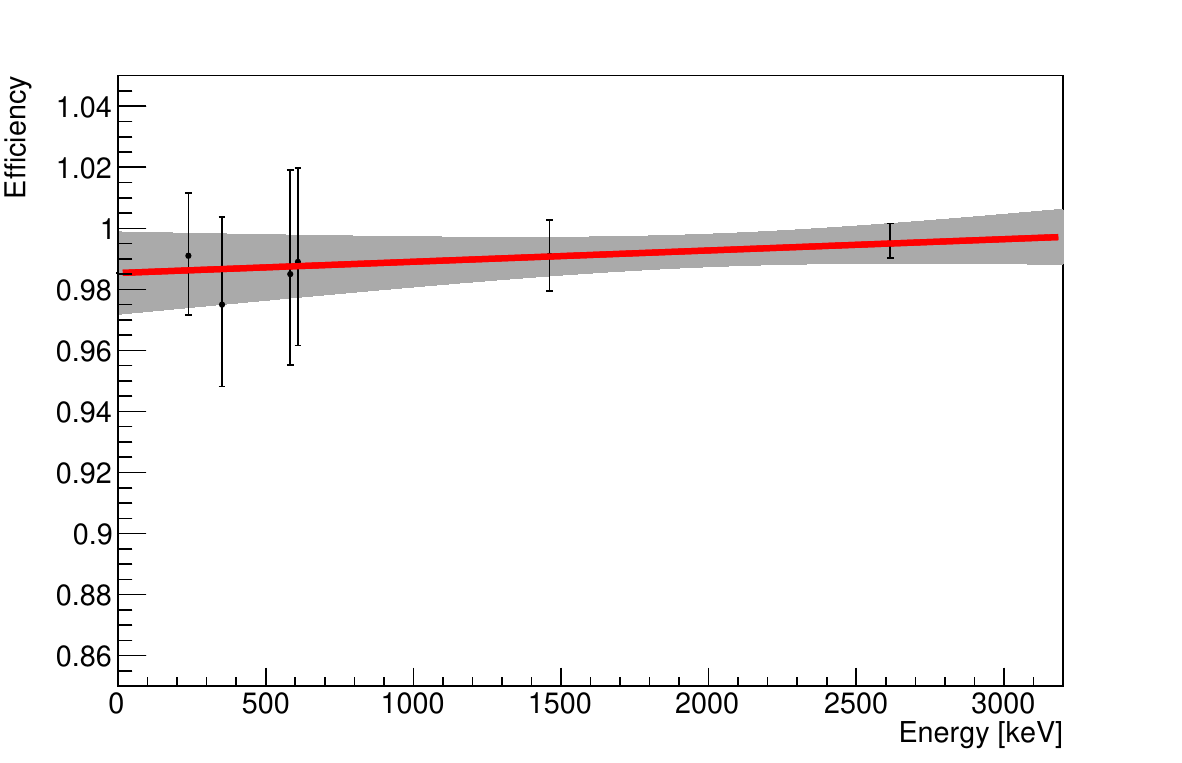}
    \caption{Plot showing the efficiency for the PCA cut ({\it Upper}) and normalized light distance cut ({\it Lower}) obtained from \mone $\gamma$ peaks as a function of the peak energy (back points). We fit these graphs to a linear polynomial (red line) and the confidence interval of this linear fit is shown in gray.}
    \label{fig:EffGraphs}
\end{figure}

We compute the efficiency of the normalized light distance cut (i.e., LY cut) and the PSD cut using a new method in this analysis. We fit the $\gamma$ peaks in the \mone data as they provide a clean sample of signal-like events, and are a more robust population with which to evaluate the efficiency, compared to using all physics events as was done previously~\cite{CuMoPRL}. In order to account for background with non-signal like events around each peak we fit the distributions of both events passing and failing each cut to a Gaussian plus linear model. The efficiency is then given as:
\begin{equation}
    \varepsilon = N_{\text{pass}}/(N_{\text{pass}} + N_{\text{fail}}).
\end{equation}
We do not expect large variation in the cut efficiency across datasets and in order to maintain sufficient statistics when using the $\gamma$ peaks we compute only the global cut efficiencies.
We estimate the uncertainty numerically by sampling from the uncertainty on the number of events in the photopeaks from the Gaussian fit. We apply the LY cut in order to gain a clean sample of events when measuring the PSD efficiency and vice-versa, which is possible due to the independence of the heat and normalized light signals. We perform this for each significant $\gamma$ peak in the \mone physics data (excluding the $^{60}$Co peaks for the LY cut as they are known to be biased due to a contaminated LD). We fit the efficiency as a function of peak energy to a linear polynomial and observe that the efficiency is consistent with being constant (between 238--2615~\,keV). We extrapolate to \qbb in order to obtain the efficiencies for each cut in order to account for any systematic energy dependence. These fits are shown in Fig.~\ref{fig:EffGraphs}.
We combine the efficiencies measured in Table~\ref{tab:effs} to determine the overall total analysis efficiency. We sample from the errors for each efficiency (assumed to be Gaussian), and obtain an estimate of the probability distribution of the total efficiency from which we extract the analysis cut efficiency with a Gaussian fit as $\varepsilon =$ (\analysiseff)\%. 
\section{Resolution Scaling and Energy Bias}
\label{sec:reso}
As there is no significant naturally occurring $\gamma$ peak near \qbb we must perform an extrapolation of the resolution as a function of energy and likewise for the energy scale bias. In order to account for variations in the performance and noise of each LMO detector over time, we obtain the energy scale extrapolations on a channel-dataset basis. Due to the excellent radiopurity and the relatively fast 2\vbb decay rate which covers most $\gamma$ peaks in the spectrum, we cannot determine this scaling directly from physics data alone. In order to have sufficient statistics, we utilize calibration data to obtain a lineshape from the 2615~\,keV $\gamma$ events which is then extrapolated to physics data.

\subsection{Resolution in Calibration Data}
\begin{figure}[htpb]
    \centering
    \includegraphics[width=\columnwidth]{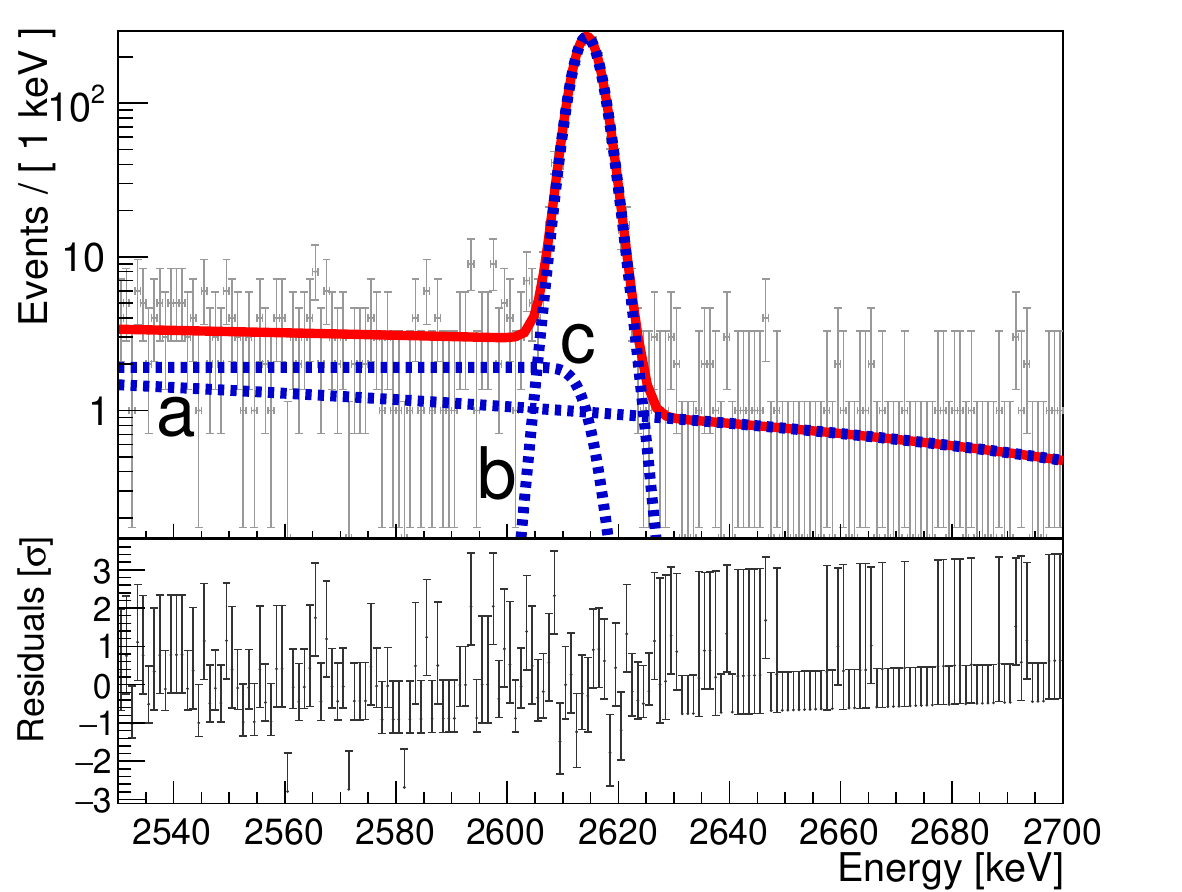}
    \caption{Simultaneous fit of calibration data 2615~\,keV $\gamma$ peak, and residuals for all channels in a single dataset with Poisson errors in each bin. Top: the summed total effective fit components (dashed blue lines) are labeled. Component (a) is the total excess background modeled by a linear fit, component (b) is the sum of Gaussian lineshapes used for each channel, and component (c) is the smeared step function to represent multi-Compton background. The simultaneous fit is shown as a solid line. Bottom: the residuals of the fit showing overall excellent agreement across the model with the central core well described.}
    \label{fig:simulfit}
\end{figure}

As in \cite{CuMoPRL} we perform a simultaneous fit of the 2615 keV peak in calibration data for each dataset. This fit is an unbinned extended maximum likelihood fit implemented using RooFit~\cite{RooFit}. We model the data in each channel as:
\begin{align}
    f_{c,d}(E) = N_c(&p_{b}\cdot f_{b}(E;l)\\&+p_{s}\cdot f_s(E;\mu_{c,d},\sigma_{c,d}) \nonumber \\&+f_{g}(E;\mu_{c,d},\sigma_{c,d})) \nonumber
\end{align}
where $c$ is the channel number, $d$ is the dataset and the functions $f_b(E;l),f_{s}(E;\mu_{c,d},\sigma_{c,d}),f_g(E;\mu_{c,d},\sigma_{c,d})$ are normalized linear background, smeared step and Gaussian functions. The parameter $l$ is the slope of the linear background, $\mu_{c,d}$ is the mean of the peak for channel $c$ in dataset $d$ and $\sigma_{c,d}$ is the corresponding standard deviation. $p_b,p_s$ are the background and smeared step ratio (these parameters are shared for all channels). $N_c$ is the number of events in the Gaussian peak, while $\sigma_c,\mu_c$ are the resolution and mean for this channel. An example of one of these fits is seen in Fig.~\ref{fig:simulfit}. We observe in each dataset that the core of the peak is well described by the model with some distortion in the low-energy tail due to the presence of pileup events due to the high event rate in calibration data. We use the individual channel-dataset widths and means in the physics data extrapolation.
\subsection{Resolution in Physics Data}
\label{sec:reso_phys}
In order to reconstruct the resolution in physics data we use a slightly different procedure compared to \cite{CuMoPRL} and \cite{CUOREPRL2020}. We fit selected peaks with the lineshape model and extract an energy dependent resolution function from this. In the previous analysis we utilized a simple Gaussian plus linear background for each peak fit on the total summed spectrum and took the ratio, $R$, of each peak resolution to the calibration summed spectrum 2615~\,keV $\gamma$ peak. Here we introduce a new exposure weighted lineshape function:
\begin{equation}
\label{eq:ls}
    f(E) = \sum_{d=1}^9 \sum_{c=1}^{19} \frac{(\exposuresymb)_{c,d}}{\exposuresymb} f_g(E;\mu,\sigma_{c,d}\cdot R),
\end{equation}
where the summation occurs over channels $c$, and datasets $d$, $\exposuresymb$ is the exposure, $f_{g}(E)$ is a Gaussian, $\mu$ is the mean of the peak and $R$ is a ratio scaling from calibration to physics data. We fit each peak in the physics data summed spectrum to this lineshape plus a linear background as a binned likelihood fit with the number of events in the peak, and the linear background, $R$ and $\mu$ as free parameters. 

\begin{figure}[htpb]
    \centering
    \includegraphics[width=\columnwidth]{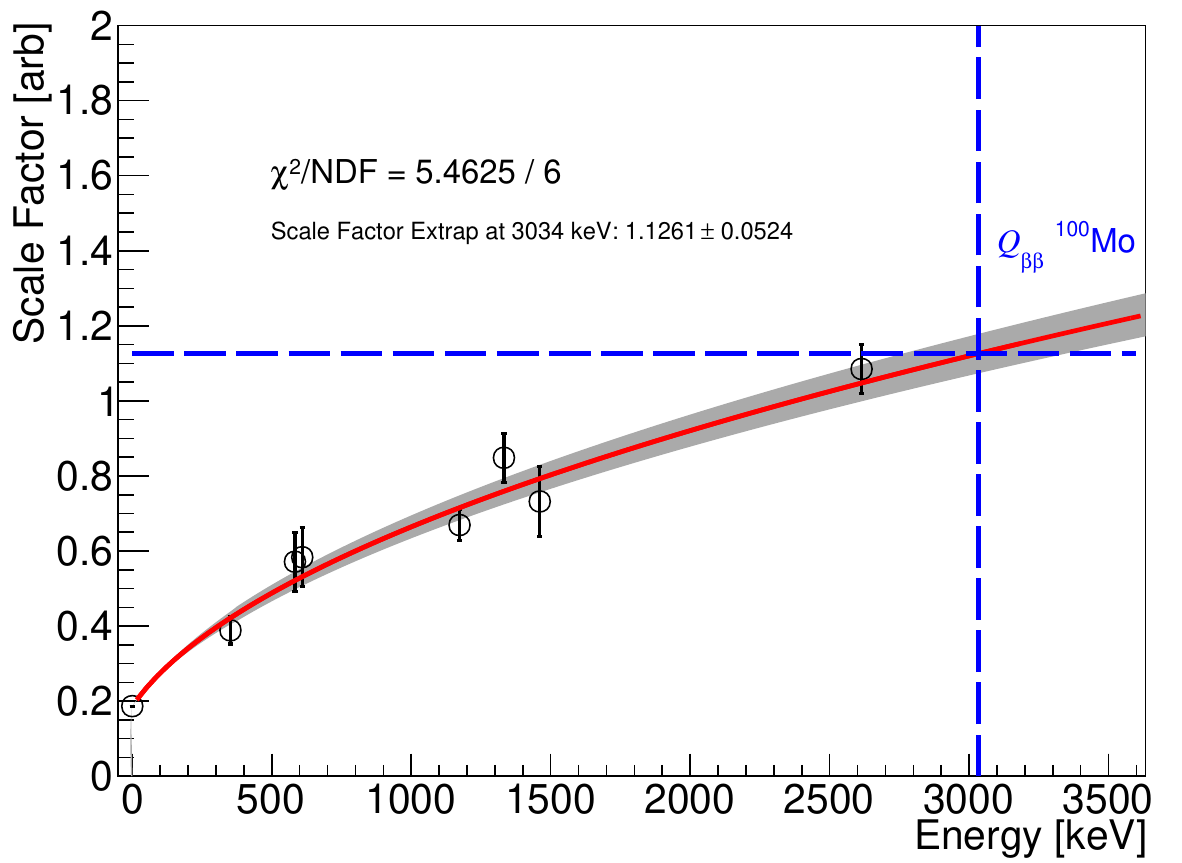}
    \caption{Resolution scaling fit showing the scaling factor $R$, between background and calibration data for each peak in physics data. We model this as $R(E) = \sqrt{p_0^2+p_1E}$ and extrapolate to \qbb to obtain a global scale factor. The choice of functional form derives from the energy resolution scaling in data following this same functional form.}
    \label{reso_scale}
\end{figure}

After all peaks in physics data have been fit we can model the resolution ratio as a function of energy. A typical functional form for the resolution of a calorimeter can be given by:
\begin{equation}
    \sigma(E)= \sqrt{\sigma_0^2+p_1E},
    \label{eq:reso_model}
\end{equation}
where the term $\sigma_0$ is related to the baseline noise in the detector, while $p_1$ characterizes any stochastic effects that degrade the resolution with increasing energy, as in~\cite{lumineu2017}. We use {\it noise} events to constrain the baseline component of the energy resolution. By fitting the distribution of noise events to the same model as the physics peaks we measure $R(0\  \mathrm{keV})$. We fit $R(E)$ for each physics peak and also the noise peak to Eq.~\ref{eq:reso_model} as shown in Fig.~\ref{reso_scale}. As in the previous analysis, we also considered a simple linear model, $\sigma = p_0+p_1E$ for the resolution scaling. Previously, in physics data there were insufficient statistics to favor one model over another, however with the additional two datasets this linear model is disfavored, as has been seen in calibration data.
Using the model in Eq.~\ref{eq:reso_model} we extrapolate the ratio at \qbb to be $R(3034~\,\mathrm{keV})=$\,\scalefactor. This number then is used to scale each of the channel-dataset dependent 2615~\,keV resolutions from the simultaneous lineshape fit in calibration data to resolutions at \qbb in physics data:
\begin{equation}
\begin{split}
   &\sigma_{c,d}(Q_{\beta\beta})= R(Q_{\beta\beta})\cdot \sigma_{c,d}\pm \\&\sqrt{ (R(Q_{\beta\beta})\cdot \sigma(\sigma_{c,d}))^2+(\sigma(R(Q_{\beta\beta}))\cdot \sigma_{c,d})^2}.
\end{split}
\end{equation}
These extrapolated resolutions are used to compute the containment efficiency (see section \ref{sec:eff}). The exposure weighted harmonic mean of the 2615~\,keV line in calibration data is $\left(6.6 \pm 0.1\right)$~\,keV FWHM. We use this to compute the effective resolution in physics data at \qbb by scaling by $R(3034~\,\mathrm{keV})$, obtaining \resolution FWHM.

\subsection{Energy Bias}
\label{sec:ebias}
\begin{figure}[htpb]
    \centering
    \includegraphics[width=\columnwidth]{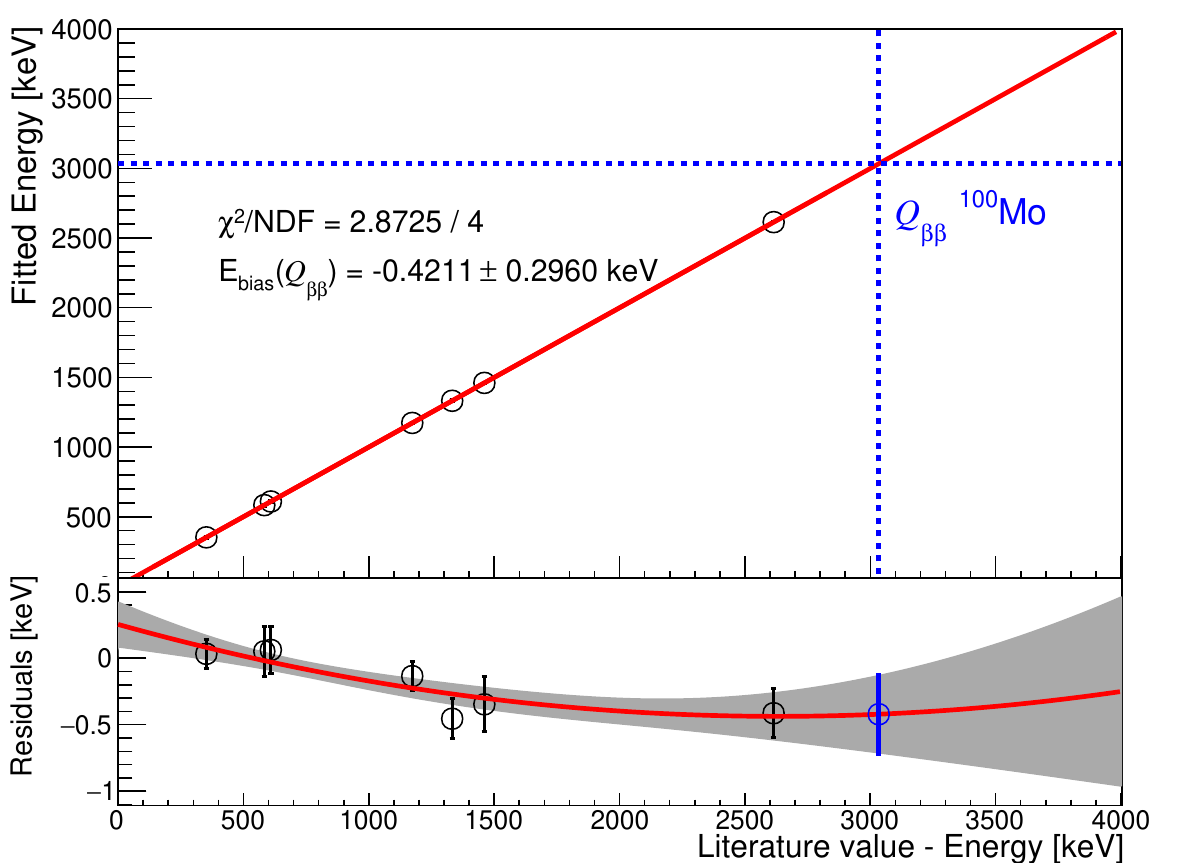}
    \caption{Energy bias in physics data. Here the best fit central value vs true value for each peak in the physics data is fit against a quadratic polynomial. The residual evaluated at \qbb is then obtained from this fit giving an estimate for the energy scale bias.}
    \label{fig:ebias}
\end{figure}

The total effective energy bias is also extracted from the fit done in physics data described in section \ref{sec:reso_phys}. Using the best fit peak locations, $\mu$ from the lineshape fit (Eq.~\ref{eq:ls}), we fit the residuals of $\mu - \mu_{\text{lit.}}$ as a function of $\mu_{\text{lit.}}$ to a second order polynomial as shown in Fig.~\ref{fig:ebias}. As in the previous analysis, we find the distribution is well described by this model and we extract the energy bias at \qbb as $E-Q_{\beta\beta} = $\,\ebias.

\section{Bayesian Fit}
\label{sec:bayes}
\subsection{Model Definition}
We use a Bayesian counting analysis to extract a limit on $T_{1/2}^{0\nu}$, similar to that in \cite{CuMoPRL}. However, due to significant improvements in the background modelling of the CUPID-Mo data we modify this analysis. We model our background in the ROI as the sum of an exponential and linear background:
\begin{equation}
    f(E) = B\cdot \Big(\frac{p_f}{\Delta E}+ (1-p_f)\cdot \frac{e^{-(E-Q_{\beta\beta})/\tau}}{N} \Big),
\end{equation}
where $B$ is the total background index (averaged over the 100 keV blinded region) in \ckky, $\Delta E$ is the width of the blinded region (100 keV), $\tau$ is the slope of the exponential and $p_f$ is the probability of flat background. Finally, $N$ is a normalization factor for the exponential. We use a counting analysis with three bins, with the expected number of counts in a bin with index $i$ given by:
\begin{align}
    \lambda_{i} = \sum_{c=1}^{19} \sum_{d=1}^9 (\exposuresymb)_{c,d}/\exposuresymb \cdot \Big( &\varepsilon_i(c,d)\cdot \Gamma^{0\nu}\frac{N_A\cdot \exposuresymb \cdot \eta}{W}\\ &+ \int_{E_{a,i}(c,d)}^{E_{b,i}(c,d)} f(E) dE \nonumber \Big).
\end{align}
The sum $c$ is over all channels and $d$ over all datasets. $\Gamma^{0\nu}$ is the $0\nu\beta\beta$ decay rate, $N_A$ is Avogadro's number, $\exposuresymb$ is the total LMO exposure, while $(\exposuresymb)_{c,d}$ is the exposure for one channel/dataset, $\eta$ is the isotopic enrichment, and $W$ is the enriched LMO molecular mass. $\varepsilon_i(c,d)$ is the total $0\nu\beta\beta$ decay detection efficiency for channel $c$, dataset $d$, and bin $i$. This is the product of the analysis efficiency (see section \ref{sec:eff}) and the containment efficiency. This is the probability for a $0\nu\beta\beta$ decay event to have energy in bin $i$ and to be \mone.
The expected number of counts is a sum of a signal contribution \mbox{$\varepsilon(c,d)\cdot N^{0\nu}$}, and a background contribution from integrating $f(E)$ between the bounds $[E_{a,i}(c,d),E_{b,i}(c,d)]$, the upper and lower bounds for the bin $i$.
The decay rate is normalized by a constant to give the number of 0\vbb decay events. The three bins used in this analysis represent lower/upper side-bands to constrain the background, and a signal region. The energy ranges of the signal region are chosen on a channel-dataset basis (see section \ref{sec:roi_opt}), and the remaining energies out of the 100 keV fit region form the sidebands. The efficiencies $\varepsilon(c,d)$ are defined for each detector-dataset from Monte Carlo (MC) simulations accounting for the energy resolution and its uncertainty. Our likelihood is then given by a binned Poisson likelihood over three bins:
\begin{equation}
    \mathcal{L} = \prod_{i=0}^2 \frac{\lambda_i^{N_i}e^{-\lambda_i}}{N_i!}.
\end{equation}
We simultaneously minimise and sample from the joint posterior distribution using the Bayesian Analysis Toolkit (BAT~\cite{BATCaldwell}). Our model parameters are:
\begin{itemize}
    \item $B$: the background index;
    \item $p_f$: the probability of flat background;
    \item $\tau$: the exponential background decay constant;
    \item $\Gamma^{0\nu}$: the $0\nu\beta\beta$ decay rate.
\end{itemize}
We also include systematic uncertainties as nuisance parameters as described in section \ref{sec:syst}.
\subsection{Optimization of the ROI}
\label{sec:roi_opt}
\begin{figure}[htpb]
    \centering
    \includegraphics[width=\columnwidth]{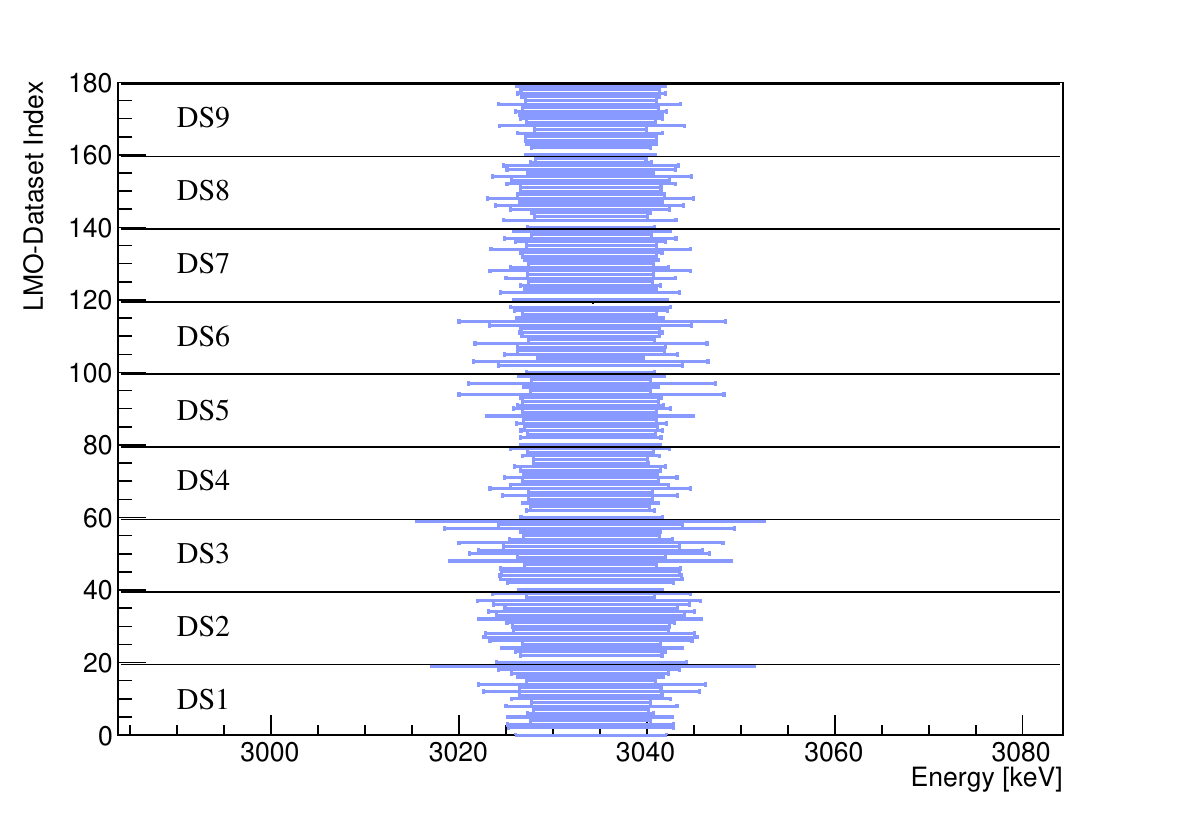}
    \caption{Summary plot of the ROI width for every channel-dataset in this analysis. Horizontal lines demarcate each of the nine separate datasets and blue lines are a particular channel's ROI. The x-axis range spans the entire 100 keV blinded region.}
    \label{fig:roi}
\end{figure}

Due to the different performance of each channel across datasets we use different ROIs for each. These are optimized using blinded data to maximize the mean expected sensitivity using the same procedure defined in \cite{CuMoPRL}. We optimize the ROI window based on the likelihood ratio defined as:
\begin{equation}
    R(c,d,E) = \frac{\mathcal{L(B)}}{\mathcal{L}(S)},
\end{equation}
where $\mathcal{L}(B)$ is the probability that an event at energy $E$ in channel $c$ and dataset $d$ is background, and $\mathcal{L}(S)$ is the same for signal. We divide the energy in 0.1~\,keV bins between 2984--3084~\,keV for each channel-dataset from which we extract the containment efficiency and estimated background. We rank these bins via the likelihood ratio:
\begin{equation}
        R(c,d,E_i) = \frac{\mathcal{L(B)}}{\mathcal{L}(S)} \propto \frac{\varepsilon_{c,d,E_i}}{B_{c,d,i}},
\end{equation}
where the background index is assumed to be constant at $5\times 10^{-3}$~\ckky (in the previous analysis we found this assumption does not significantly impact the results~\cite{CuMoPRL}). We then optimize the choice of the maximum allowed likelihood ratio to include by maximizing the mean limit setting sensitivity, as a Poisson counting analysis:
\begin{equation}
    S=\sum_{n=0}^3 p(n)\cdot S(n),
\end{equation}
with the limit, $S(n)$, of 2.3 counts in the case of zero events, 3.9 for one event, etc., and $p(n)$ is the probability of observing $n$ counts based on the expected background rate. The chosen channel-dataset based ROIs are shown in Fig.~\ref{fig:roi}, with an exposure weighted effective ROI width of \roiWidth, corresponding to ($2.3\pm 0.6$) FWHM at \qbb.

\subsection{Containment Efficiency}
Once the channel-dataset based ROIs have been chosen we can compute the containment efficiency for each channel and dataset pair. This efficiency is evaluated using Geant4 MC simulations, accounting for the energy resolutions extracted in section \ref{sec:reso}. The average containment efficiency is ($75.9 \pm 1.1$)\%.
To estimate the systematic uncertainty from the MC simulations we vary the simulated crystal dimensions and Geant4 production cuts resulting in a $1.5\%$ relative uncertainty.

\subsection{Extraction of the Background Prior}
The most significant prior probabilities in our analysis are for the signal rate $\Gamma^{0\nu}$ and the background index $B$. Due to the very low \cumo backgrounds and a relatively small exposure, data around the ROI does not constrain $B$ well. However, detailed Geant4 modelling does provide a measurement of the background averaged over our 100~\,keV blinded region (a forthcoming publication on the background modelling is in preparation). This fit models our experimental data in bin $i$ as (in units of counts/keV):
\begin{equation}
    \mu_i = \sum_{j=1}^k N^{\text{MC}}_{j,i}\cdot f_j/\Delta E,
\end{equation}
where the sum is over all simulated MC contributions, $N^{\text{MC}}_{j,i}$ is the number of events in the simulated MC spectra $j$ and bin $i$, and $f_j$ is a factor we obtain from the fit. This fit is performed using a Bayesian fit based on JAGS~\cite{JAGS_Proceeding,JAGS3}, similar to~\cite{CUORE.2nu, CUPID0.BM}. It estimates the joint posterior distribution of the parameters $f_j$, and we sample from this distribution at each step in the Markov chain computing:
\begin{equation}
    B_i = \sum_{j=1}^{k}\frac{N^{\text{MC}}_{j,i}\cdot f_{j}}{\exposuresymb }.
\end{equation}
From the marginalized posterior distribution of the observable background index we obtain:
\begin{equation}
    \bi.
\end{equation}
This value is used as a prior in our Bayesian fit with a split-Gaussian distribution; two Gaussian distributions with the same mode are combined such that values on either side of the mode have different variances. We have found that in the case of observing zero events, this prior does not change the observed limit. However, if some events are observed, this is a more conservative choice than a non-informative flat prior since it prevents the background index from floating to high values that are strongly disfavored by the background model.

\begin{figure}[htbp]
\centering
    \includegraphics[width=\columnwidth]{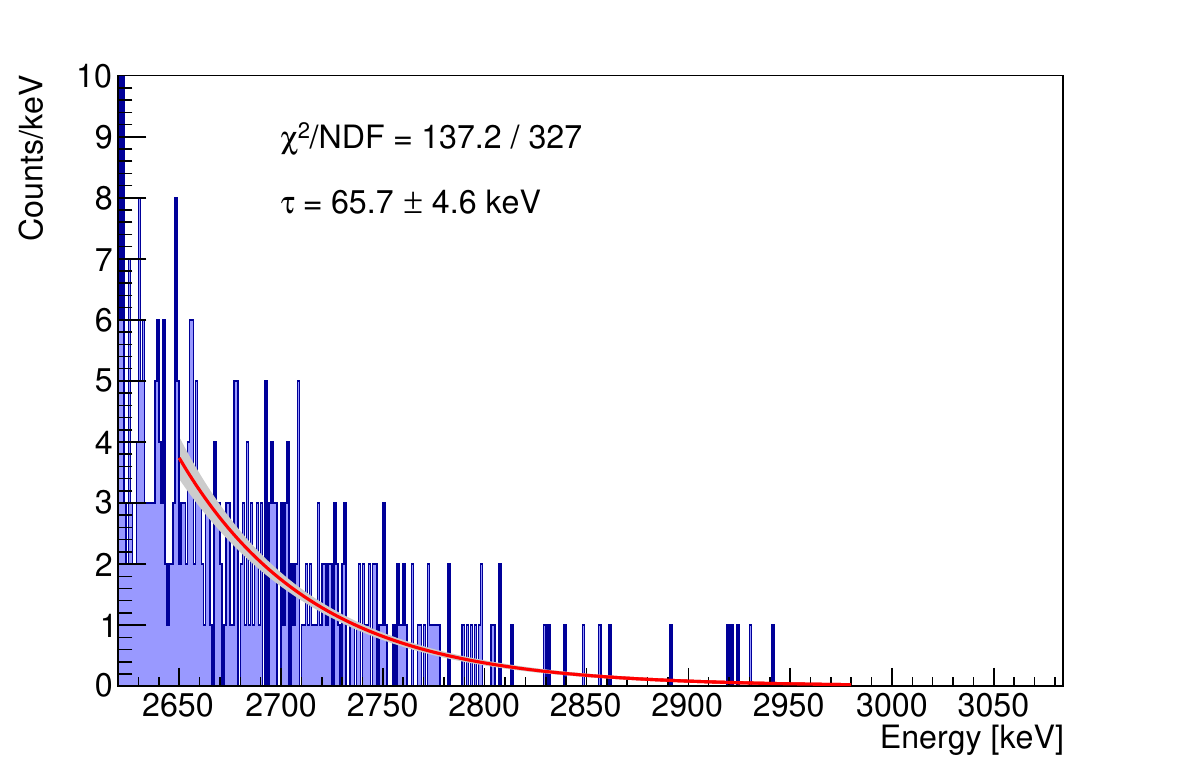}
    \caption{The exponential plus flat background fit used to determine a prior on the slope of any exponential background. The fit favors an exponential term with no flat background, however owing to the low statistics it is not sufficient to rule out the presence of a flat background term.}
    \label{fig:tauexp}
\end{figure}

To extract a prior on the slope of the exponential background, $\tau$, we perform a fit to the blinded data (between $2650$ to $2980$ keV) to a constant plus exponential model, as seen in Fig.~\ref{fig:tauexp}. This results in a best fit of $\tau  = \left(65.7\pm 4.6\right) \ \mathrm{keV}$, which is used as a prior in our analysis. The probability of the background being uniform (instead of exponential) is given a uniform prior between $[0,1]$.
\subsection{Systematic Uncertainties}
\label{sec:syst}
We include systematic uncertainties in our Bayesian fit as nuisance parameters, in particular we account for uncertainties in:
\begin{itemize}
    \item cut efficiencies;
    \item isotopic enrichment;
    \item containment efficiency.
\end{itemize}
These are each given Gaussian prior distribution with the values from sections \ref{sec:eff} and \ref{sec:reso} as indicated in Table~\ref{tab:nuisance}.
\begin{table*}[hptb]
    \centering
    \caption{Nuisance parameters for the Bayesian model and their central values and prior type. The background index has an asymmetric uncertainty and is treated as a split-Gaussian, with each side corresponding to the different asymmetric uncertainty values. The signal rate is treated as a uniform prior in the positive domain.}
    \begin{tabular}{lcc}
    \centering
    Nuisance Parameter & Value & Prior type \\ \hline \hline
    Rate & [0, $40\times 10^{-24}$]~\,yr$^{-1}$ & Uniform \\
    Isotopic Enrichment & $0.966 \pm 0.002$ & Gauss. \\
    0\vbb decay containment (MC) & $1.000 \pm 0.015$ & Gauss. \\
    Analysis Efficiency (global) & $0.8843 \pm 0.0180$ & Gauss. \\
    Background Index & $(4.7^{+1.7}_{-1.7}) \times 10^{-3}$ \ckky & Split Gauss. \\
    Probability of flat background ($p$) & [0, 1] & Uniform \\ 
    Exponential background slope ($\tau$)& $\left(65.7 \pm 4.6\right)$~\,keV & Gauss.
    \end{tabular}
    \label{tab:nuisance}
\end{table*}
As in \cite{CuMoPRL} these uncertainties are marginalized over and are automatically included in our limit. We note that the systematic uncertainties from the energy bias and resolution scaling are incorporated in the computation of the containment efficiency. We chose a uniform prior on the rate, $\Gamma^{0\nu} \in [0, 40\times 10^{-24}$]~\,yr$^{-1}$. This is consistent with the standard practice for 0\vbb decay analysis \cite{GERDA2020,EXO200.2019,CUORE1ton}. The range is large enough that it has minimal impact on the possible result, and provides as little information as possible on the rate to avoid possible bias.
\section{Results}
\label{sec:results}
\begin{figure}[htpb]
    \centering
    \includegraphics[width=\columnwidth]{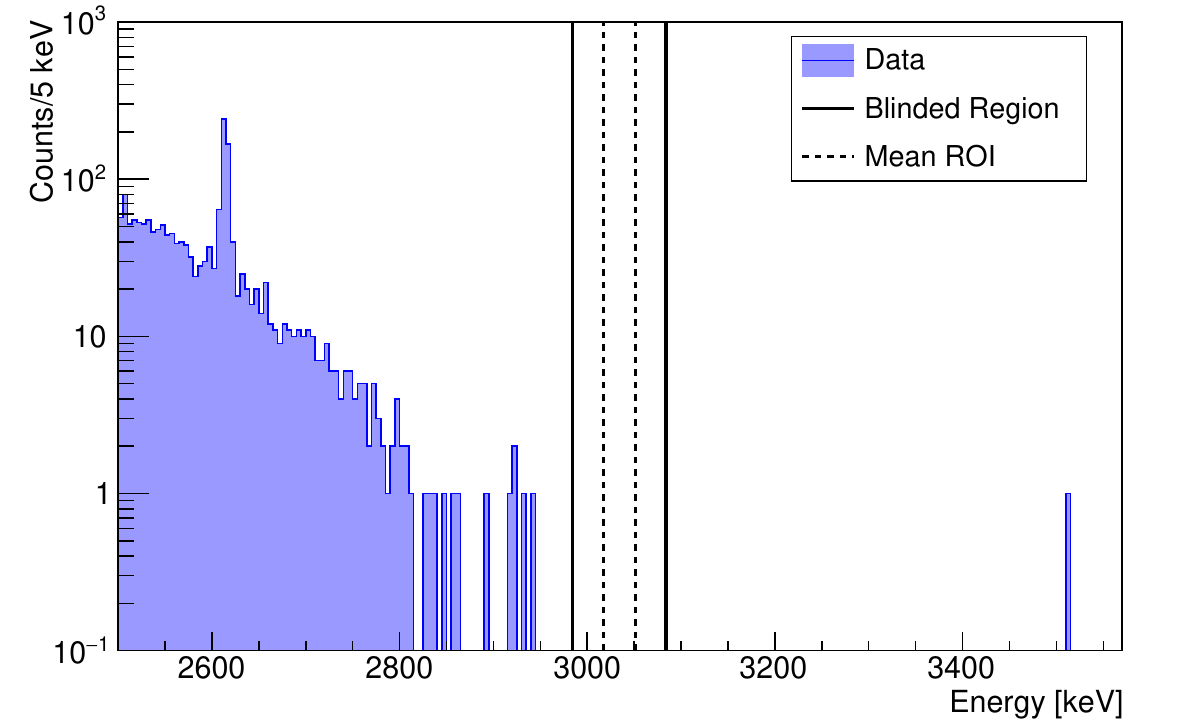}
    \caption{The unblinded background spectrum near the ROI for \exposure~\,\kgyr of data (\isoexposure~\,\kgyr for \isomo). After application of all cuts we observe no events in both the ROI and in the full 100~\,keV blinded region. In this work, the event near 3200~\,keV present in the previous analysis~\cite{CuMoPRL}, was tagged as coincident with the improved muon veto. The exposure weighted mean ROI (\roiMedian) is shown with dashed lines, and the full blinded region is within the solid lines.}
    \label{fig:unblindedROI}
\end{figure}
\begin{figure}[htbp]
    \centering
    \includegraphics[width=\columnwidth]{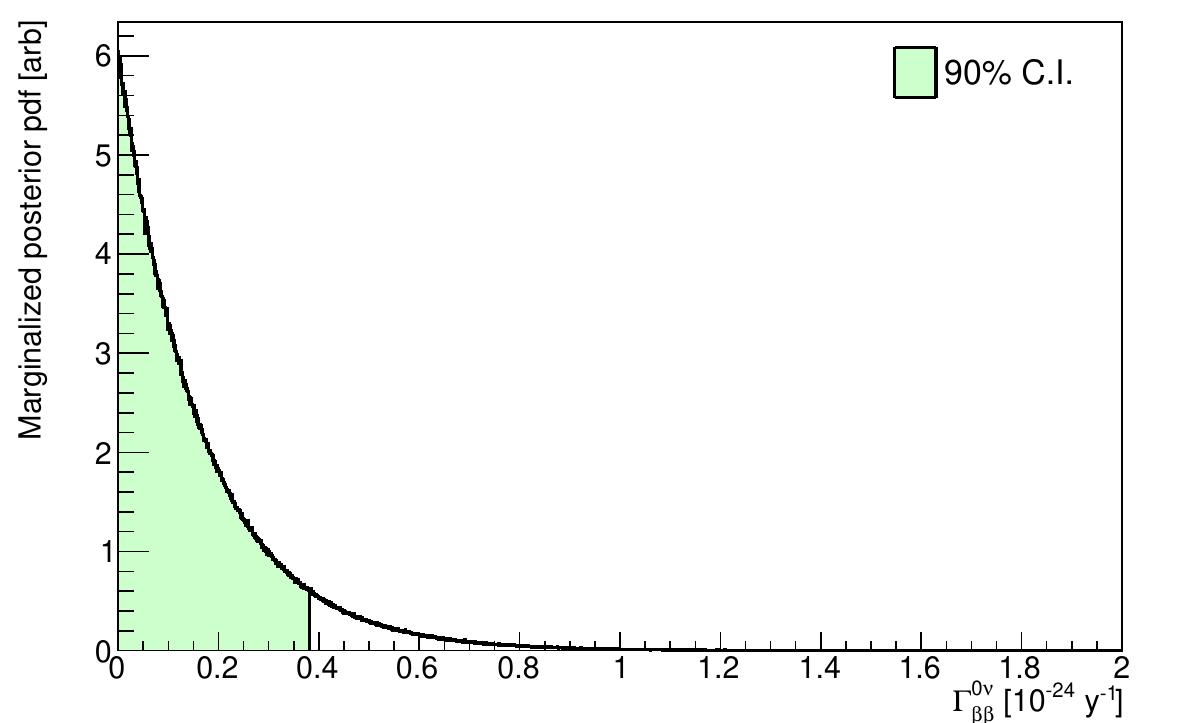}
    \caption{The posterior distribution of the decay rate for 0\vbb decay run with all nuisance parameters floating. The shaded area under the curve represents the 90\% C.I. with upper limit at a value of $\Gamma^{0\nu}$ = \rateNumber.}
    \label{fig:rposterior}
\end{figure}

After unblinding our data, we observe zero events in the channel-dataset ROIs and zero events in the side-bands, as shown in Fig.~\ref{fig:unblindedROI}. This leads to an upper limit on the decay rate $\Gamma^{0\nu}$ including all systematics of:
\begin{equation}
    \rate
\end{equation}
or:
\begin{equation}
    \hlifeResultEqn
\end{equation}
This limit surpasses our first result of $T^{0\nu}_{1/2} > 1.5 \times 10^{24}$~\,yr~\cite{CuMoPRL}, becoming a new leading limit on 0\vbb decay in \isomo. The posterior distribution of the decay rate is shown in Fig.~\ref{fig:rposterior}. We find that this can be fit well by a single exponential as expected for a background-free measurement. We extract:
\begin{equation}
    p(\Gamma^{0\nu}|D_{\text{CUPID-Mo}}) = \lambda\cdot e^{-\lambda \cdot \Gamma^{0\nu}}d\Gamma^{0\nu},
\end{equation}
where
\begin{equation}
    \posteriorexp,
\end{equation}
and $D_{\text{CUPID-Mo}}$ is the CUPID-Mo data.
We can extract the 90\% C.I. on the signal counts from the posterior, resulting in an upper limit of $S < 2.3$ counts (90\% C.I.), consistent with what one would expect from a Poisson counting experiment with zero observed events. Our Bayesian analysis leads to a non-zero background index in the 100~\,keV fit region with a 1$\sigma$ interval of:
\begin{equation}
    \bipost.
\end{equation}
This is mostly consistent with the informative background model prior. Further studies are ongoing to include extra information into the background model fit (i.e. constraints on pileup from simulation or calibration data) to reduce this uncertainty. The posterior distributions for the exponential background parameters are consistent with the priors derived from the fit of the 2\vbb decay spectrum in an energy interval between 2650$-$2980~\,keV (as done previously~\cite{CuMoPRL}).
\begin{figure}[htbp]
    \centering
    \includegraphics[width=0.99\columnwidth]{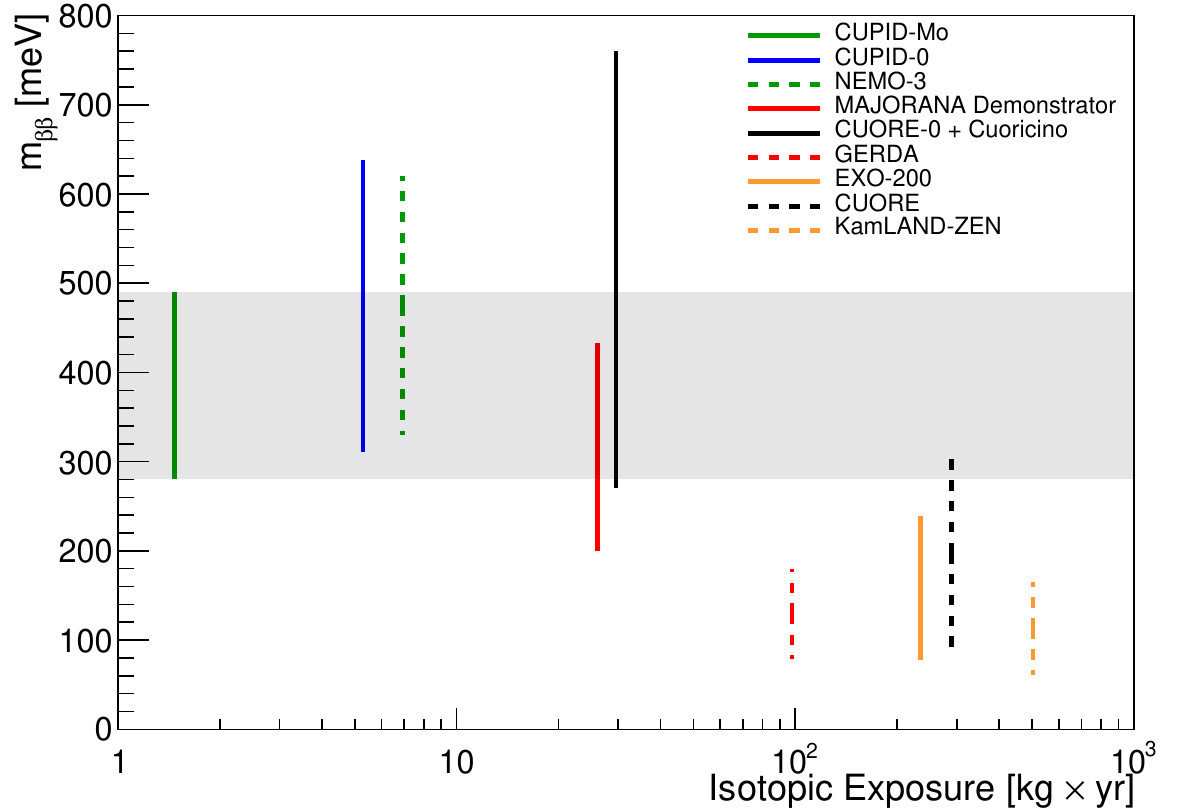}
    \caption{Published \mbbavg values as a function of isotopic exposure for several experiments~\cite{Q0FinalPRC, CUPID0Final, NEMO30vMo, MJD2019, EXO200.2019, GERDA2020, CUORE1ton, KamLANDZen2016}. While each experiment has utilized different sets of nuclear matrix elements, the published results represent a standard set of values we can compare this work's result to. The spread for \cumo results is shown with a band for illustrative purposes only, indicating the \mbbavg range as it relates to other experiments more clearly and showing the promise of this technology with only a relatively modest isotopic exposure.}
    \label{fig:mbbPlot}
\end{figure}
In order to study the effect of systematics we perform a series of fits allowing only one nuisance parameter to float at a time, with all others fixed to their prior's central value. The nuisance parameters we allow to float are the isotopic abundance, MC containment efficiency factor, and analysis efficiency. These are compared against fits with all parameters fixed (e.g., a statistics-only run), and again allowing all parameters to float. For each category of test we run $\sim$1000 toys, each generating $10^{4}$ Markov chains. We find that relative to statistics-only runs (i.e., fixing all nuisance parameters), the effect of each nuisance parameter on the marginalized rate is less than 1\%. The largest impact originates from the global analysis efficiency at $\sim$ 0.7\%. This is not surprising as the relative uncertainty on the analysis efficiency is high compared to the other parameters.

We interpret the obtained half-life limit on the 0\vbb decay in \isomo in the framework of light Majorana neutrino exchange. We utilize $g_{A} = 1.27$, and phase space factors from~\cite{Kotila2012, Mirea2015}. We consider various nuclear matrix elements from~\cite{NME.Rath13, NME.Simkovic13, NME.Vaquero13, NME.Barea15, NME.Pirinen15, NME.Song17, NME.Simkovic18, NME.Rath19}. This results in a limit on the effective Majorana neutrino mass of:
\begin{equation}
    \mbbResultEqn.
\end{equation}
This result improves upon the previous constraint by virtue of an increased \isomo exposure in the new processing and is set with a very modest exposure of \isoexposure\kgyr of $^{100}$Mo. This is seen in Fig.~\ref{fig:mbbPlot} which shows this result in the context of other experiments, indicating the promise of utilizing \isomo as a 0\vbb decay search isotope. 
\section{Conclusions}
\label{sec:conc}

In this work, we implemented refined data production and analysis techniques with respect to the previous result~\cite{CuMoPRL}. We report a final 0\vbb decay half-life limit of \hlifeResult with a relatively modest exposure of \exposure \kgyr (\isoexposure \kgyr in \isomo), with a resulting limit on the effective Majorana mass of \mbbResult. We show that an iterative channel-channel time offset correction is feasible and significantly improves the ability to tag multiple crystal events while reducing accidental coincidences. This results in a highly efficient single-scatter cut, and a more pure higher multiplicity spectra, which is useful for analyses such as decay to excited states and the development of a background model. We have also shown an improved method used for particle identification by utilizing normalized light energy quantities derived from the absolute LD calibration. This allows for an improvement in the rejection of $\alpha$ events with a high efficiency and relatively conservative cut. The pulse shape discrimination is improved via a cleaner training sample, run-by-run normalization and full energy dependence correction. It is further enhanced by combination of pulse shape parameters derived from the optimally filtered waveform. Further improvements may be possible with better tuned pulse templates and a multivariate discrimination using portions of the waveform to allow for even more pileup rejection. Finally, the very low contamination of the LMO detectors also allows for the implementation of extended delayed coincidence cuts to reject not just $^{212}$Bi-$^{208}$Tl decay chain events, but also $^{222}$Rn-$^{214}$Bi and $^{218}$Po-$^{214}$Bi decay chain events, allowing for the reduction of the background in the high energy region. This type of cut in particular may be especially useful for a larger scale experiment such as CUPID~\cite{CUPID:2019} due to the ability to remove potentially dangerous $\beta$ events.

The result of these enhanced analysis steps produces a total analysis efficiency of (\analysiseff)\% or combining with the containment efficiency, a total 0\vbb decay efficiency of (\totaleff)\%. This high total efficiency, along with low background index, and excellent energy resolution at \qbb of \resolution FWHM show that the potential for scintillating \enrLMO crystals coupled to complimentary LDs in a larger experiment such as CUPID is entirely feasible. Analysis techniques developed here can be easily applied to larger datasets.

The \cumo data can be used to extract other physics results. The analysis techniques described here have been used for an analysis of decays to excited states (publication forthcoming). Other foreseen analyses include spin-dependent low-mass dark matter searches via interaction with $^{7}$Li~\cite{Li.DM, helisThesis} in the \lmo and axion searches~\cite{Li.Axion}. \cumo has succeeded in demonstrating the feasibility of scintillating calorimeters for use in 0\vbb decay searches, having demonstrated that backgrounds from $\alpha$'s can be easily rejected via scintillation light, and that pulse-shape rejection techniques can be utilized with high efficiency.

\section{Acknowledgements}
This work has been partially performed in the framework of the LUMINEU program, a project funded by the Agence Nationale de la Recherche (ANR, France). The help of the technical staff of the Laboratoire Souterrain de Modane and of the other participant laboratories is gratefully acknowledged. 
We thank the mechanical workshops of CEA/SPEC for their valuable contribution in the detector conception and of LAL (now IJCLab) for the detector holders fabrication. 
F.A. Danevich, V.V. Kobychev, V.I. Tretyak and M.M. Zarytskyy were supported in part by the National Research Foundation of Ukraine Grant No. 2020.02/0011. O.G. Polischuk was supported in part by the project “Investigations of rare nuclear processes” of the program of the National Academy of Sciences of Ukraine “Laboratory of young scientists”. A.S. Barabash, S.I. Konovalov, I.M. Makarov, V.N. Shlegel and V.I. Umatov were supported by the Russian Science Foundation under grant No. 18-12-00003. We acknowledge the support of the P2IO LabEx (ANR-10-LABX0038) in the framework “Investissements d’Avenir” (ANR-11-IDEX-0003-01 – Project “BSM-nu”) managed by the Agence Nationale de la Recherche (ANR), France. 
Additionally the work is supported by the Istituto Nazionale di Fisica Nucleare (INFN) and by the EU Horizon2020 research and innovation program under the Marie Sklodowska-Curie Grant Agreement No. 754496. This work is also based on support by the US Department of Energy (DOE) Office of Science under Contract Nos. DE-AC02-05CH11231, and by the DOE Office of Science, Office of Nuclear Physics under Contract Nos. DE-FG02-08ER41551, DE-SC0011091; by the France-Berkeley Fund, the MISTI-France fund and  by the Chateau-briand Fellowship of the Office for Science \& Technology of the Embassy of France in the United States. This research used resources of the National Energy Research Scientific Computing Center (NERSC).
This work makes use of the DIANA data analysis software which has been developed by the CUORICINO, CUORE, LUCIFER, and CUPID-0 Collaborations. 

\noindent
Russian and Ukrainian scientists have given and give crucial contributions to CUPID-Mo. For this reason, the CUPID-Mo collaboration is particularly sensitive to the current situation in Ukraine. The position of the collaboration leadership on this matter, approved by majority, is expressed at \url{https://cupid-mo.mit.edu/collaboration#statement}. Majority of the work described here was completed before February 24, 2022. 

\bibliographystyle{spphys}       
\bibliography{biblio}   

\begin{thebibliography}{10}
\providecommand{\url}[1]{{#1}}
\providecommand{\urlprefix}{URL }
\expandafter\ifx\csname urlstyle\endcsname\relax
  \providecommand{\doi}[1]{DOI \discretionary{}{}{}#1}\else
  \providecommand{\doi}{DOI \discretionary{}{}{}\begingroup
  \urlstyle{rm}\Url}\fi

\bibitem{PhysRevLett.81.1562}
Y.~Fukuda, et~al., Phys. Rev. Lett. \textbf{81}, 1562 (1998)

\bibitem{Ahmad2001}
Q.R. Ahmad, et~al., Phys.~Rev.~Lett. \textbf{87}, 071301 (2001)

\bibitem{MajoranaPaper}
E.~Majorana, Il Nuovo Cimento \textbf{14}(4), 171 (1937)

\bibitem{PhysRevD.22.2227}
J.~Schechter, J.W.F. Valle, Phys. Rev. D \textbf{22}, 2227 (1980)

\bibitem{Racah1938}
G.~Racah, Il Nuovo Cimento \textbf{14}(7), 322 (1937)

\bibitem{Pontecorvo1968}
B.~Pontecorvo, Sov.~Phys.~JETP \textbf{26}, 984 (1968)

\bibitem{FUKUGITA198645}
M.~Fukugita, T.~Yanagida, Phys.~Lett.~B \textbf{174}(1), 45  (1986)

\bibitem{DAVIDSON2008105}
S.~Davidson, E.~Nardi, Y.~Nir, Physics Reports \textbf{466}(4), 105 (2008)

\bibitem{Furry1939}
W.H. Furry, Phys. Rev. \textbf{56}, 1184 (1939)

\bibitem{PhysRevD.25.2951}
J.~Schechter, J.W.F. Valle, Phys. Rev. D \textbf{25}, 2951 (1982)

\bibitem{Bilenky2015}
S.M. Bilenky, C.~Giunti, International Journal of Modern Physics A
  \textbf{30}(04n05), 1530001 (2015)

\bibitem{Goswami2015}
S.~Dell'Oro, S.~Marcocci, M.~Viel, F.~Vissani, Advances in High Energy Physics
  \textbf{2016}, 2162659 (2016)

\bibitem{Dolinski2019}
M.J. Dolinski, A.W. Poon, W.~Rodejohann, Annual Review of Nuclear and Particle
  Science \textbf{69}, 219 (2019)

\bibitem{GERDA2020}
M.~Agostini, et~al., Phys. Rev. Lett. \textbf{125}, 252502 (2020)

\bibitem{MJD2019}
S.I. Alvis, et~al., Phys. Rev. C \textbf{100}, 025501 (2019)

\bibitem{CUPID0Final}
O.~Azzolini, et~al., Phys. Rev. Lett. \textbf{123}, 032501 (2019)

\bibitem{CUORE1ton}
D.Q. Adams, et~al., Nature \textbf{604}, 53 (2022)

\bibitem{CUOREPRL2020}
D.Q. Adams, et~al., Phys. Rev. Lett. \textbf{124}, 122501 (2020)

\bibitem{CuMoPRL}
E.~Armengaud, et~al., Phys. Rev. Lett. \textbf{126}(18), 181802 (2021)

\bibitem{NEMO30vMo}
R.~Arnold, et~al., Phys. Rev. D \textbf{92}, 072011 (2015)

\bibitem{KamLANDZen2016}
A.~Gando, et~al., Phys.~Rev.~Lett. \textbf{117}, 082503 (2016)

\bibitem{EXO200.2019}
G.~Anton, et~al., Phys. Rev. Lett. \textbf{123}, 161802 (2019)

\bibitem{Deppisch2012}
F.F. Deppisch, M.~Hirsch, H.~P{\"a}s, J. Phys G: Nucl. Part. Phys.
  \textbf{39}(12), 124007 (2012)

\bibitem{Rodejohann2012}
W.~Rodejohann, J. Phys G: Nucl. Part. Phys. \textbf{39}(12), 124008 (2012)

\bibitem{PhysRevD.68.034016}
G.~Pr\'ezeau, M.~Ramsey-Musolf, P.~Vogel, Phys. Rev. D \textbf{68}, 034016
  (2003)

\bibitem{Atre2009}
A.~Atre, T.~Han, S.~Pascoli, B.~Zhang, Journal of High Energy Physics
  \textbf{2009}(05), 030 (2009)

\bibitem{Blennow2010}
M.~Blennow, E.~Fernandez-Martinez, J.~Lopez-Pavon, J.~Men{\'e}ndez, Journal of
  High Energy Physics \textbf{2010}(7), 96 (2010)

\bibitem{MITRA201226}
M.~Mitra, G.~Senjanovi{\'c}, F.~Vissani, Nuclear Physics B \textbf{856}(1), 26
  (2012)

\bibitem{Cirigliano2018}
V.~Cirigliano, W.~Dekens, J.~de~Vries, M.L. Graesser, E.~Mereghetti, Journal of
  High Energy Physics \textbf{2018}(12), 97 (2018)

\bibitem{TRETYAK200283}
V.I. Tretyak, Y.G. Zdesenko, Atomic Data and Nuclear Data Tables
  \textbf{80}(1), 83 (2002)

\bibitem{lumineu2017}
E.~Armengaud, et~al., Eur. Phys. J. C \textbf{77}(11), 785 (2017)

\bibitem{AMORE2019}
V.~Alenkov, et~al., Eur. Phys. J. C \textbf{79}(9), 791 (2019)

\bibitem{DPoda2021}
D.~Poda, Physics \textbf{3}(3), 473 (2021)

\bibitem{PirroScintBolo}
S.~Pirro, J.W. Beeman, S.~Capelli, M.~Pavan, E.~Previtali, P.~Gorla, Phys.
  Atom. Nucl. \textbf{69}(12), 2109 (2006)

\bibitem{Tabarelli2009}
T.~Tabarelli~de Fatis, Eur. Phys. J. C \textbf{65}(1), 359 (2009)

\bibitem{Cardani2013LMO}
L.~Cardani, et~al., JINST \textbf{8}(10), P10002 (2013)

\bibitem{Giuliani2018}
A.~Giuliani, F.~Danevich, V.~Tretyak, Eur. Phys. J. C \textbf{78}, 272 (2018)

\bibitem{sym13122255}
A.~Zolotarova, Symmetry \textbf{13}(12), 2255 (2021)

\bibitem{CUPID:2019}
W.R. Armstrong, et~al., arXiv 1907.09376  (2019)

\bibitem{CUORE_cryostat}
C.~Alduino, et~al., Cryogenics \textbf{102}, 9  (2019)

\bibitem{CuMo_instrument}
E.~Armengaud, et~al., Eur. Phys. J. C \textbf{80}(1), 44 (2020)

\bibitem{Poda2017}
D.V. Poda, et~al., AIP Conference Proceedings \textbf{1894}, 020017 (2017)

\bibitem{Haller1984}
E.E. Haller, N.P. Palaio, M.~Rodder, W.L. Hansen, E.~Kreysa, in \emph{{Neutron
  Transmutation Doping of Semiconductor Materials}}, ed. by R.D. Larrabee
  (Springer US, Boston, MA, 1984), pp. 21--36

\bibitem{Armengaud_2017}
E.~Armengaud, et~al., JINST \textbf{12}(08), P08010 (2017)

\bibitem{PhysRevC.93.045503}
C.~Alduino, et~al., Phys. Rev. C \textbf{93}, 045503 (2016)

\bibitem{Apollo_Domizio_2018}
S.D. Domizio, et~al., JINST \textbf{13}(12), P12003 (2018)

\bibitem{OT_Domizio_2011}
S.D. Domizio, F.~Orio, M.~Vignati, JINST \textbf{6}(02), P02007 (2011)

\bibitem{Alessandrello1998}
A.~Alessandrello, et~al., Nucl. Instr. Meth. A \textbf{412}(2), 454 (1998)

\bibitem{Oreglia1980}
M.~Oreglia, A study of the reactions $\psi'\rightarrow{\gamma\gamma\psi}$.
\newblock Ph.D. thesis, SLAC (1980)

\bibitem{PCA_Huang_2021}
R.~Huang, et~al., JINST \textbf{16}(03), P03032 (2021)

\bibitem{Huang2021}
R.~Huang, Searching for 0$\nu\beta\beta$ decay with {CUORE} and {CUPID}.
\newblock Ph.D. thesis, UC Berkeley (2021)

\bibitem{CowanMetric}
G.~Cowan, K.~Cranmer, E.~Gross, O.~Vitells, Eur. Phys. J. C \textbf{71}(2),
  1554 (2011)

\bibitem{CUOREPRL2017}
C.~Alduino, et~al., Phys. Rev. Lett. \textbf{120}, 132501 (2018)

\bibitem{PhysRevC.104.015501}
A.~Armatol, et~al., Phys. Rev. C \textbf{104}, 015501 (2021)

\bibitem{BandacBetaDepth}
I.C. Bandac, et~al., Applied Physics Letters \textbf{118}(18), 184105 (2021)

\bibitem{CUPIDMo.2nu.precise}
E.~Armengaud, et~al., Eur. Phys. J. C \textbf{80}(7), 674 (2020)

\bibitem{Denys.Andrea.LTD}
D.~Poda, A.~Giuliani, International Journal of Modern Physics A
  \textbf{32}(30), 1743012 (2017)

\bibitem{CROSS}
I.C. Bandac, et~al., Journal of High Energy Physics \textbf{2020}(1), 18 (2020)

\bibitem{Nones2021}
C.~Nones, in \emph{Talk given at the 17$^{th}$ Int. Conf. on Topics in
  Astroparticle Underground Physics (TAUP 2021), online} (2021).
\newblock
  \urlprefix\url{https://indico.ific.uv.es/event/6178/contributions/15716/}

\bibitem{Schmidt2020}
B.~Schmidt, et~al., Journal of Physics: Conference Series \textbf{1468}(1),
  012129 (2020)

\bibitem{Poda2020}
D.V. Poda, in \emph{Poster Presented at the XXIX International Conf. on
  Neutrino Physics and Astrophysics} (2020).
\newblock
  \urlprefix\url{https://indico.fnal.gov/event/19348/contributions/186385/}

\bibitem{SCHMIDT201328}
B.~Schmidt, et~al., Astroparticle Physics \textbf{44}, 28 (2013)

\bibitem{helisThesis}
D.L. Helis, {Searching for neutrinoless double-beta decay with scintillating
  bolometers}.
\newblock Theses, {Universit{\'e} Paris-Saclay} (2021).
\newblock \urlprefix\url{https://tel.archives-ouvertes.fr/tel-03442659}

\bibitem{RooFit}
W.~{Verkerke}, D.~{Kirkby}, arXiv:physics/0306116  (2003)

\bibitem{BATCaldwell}
A.~Caldwell, D.~Koll{\'a}r, K.~Kr{\"o}ninger, Computer Physics Communications
  \textbf{180}(11), 2197 (2009)

\bibitem{JAGS_Proceeding}
M.~Plummer, 3rd International Workshop on Distributed Statistical Computing
  (DSC 2003) \textbf{124} (2003)

\bibitem{JAGS3}
D.~Chiesa, E.~Previtali, M.~Sisti, Ann.~Nucl.~Energy \textbf{70}, 157  (2014)

\bibitem{CUORE.2nu}
D.Q. Adams, et~al., Phys. Rev. Lett. \textbf{126}, 171801 (2021)

\bibitem{CUPID0.BM}
O.~Azzolini, et~al., Eur.~Phys.~J.~C \textbf{79}(7), 583 (2019)

\bibitem{Q0FinalPRC}
C.~Alduino, et~al., Phys.~Rev.~C \textbf{93}(4), 045503 (2016)

\bibitem{Kotila2012}
J.~Kotila, F.~Iachello, Phys.~Rev.~C \textbf{85}, 034316 (2012)

\bibitem{Mirea2015}
M.~Mirea, T.~Pahomi, S.~Stoica, Romanian Reports in Physics \textbf{67}, 872
  (2015)

\bibitem{NME.Rath13}
P.K. Rath, R.~Chandra, K.~Chaturvedi, P.~Lohani, P.K. Raina, J.G. Hirsch, Phys.
  Rev. C \textbf{88}, 064322 (2013)

\bibitem{NME.Simkovic13}
F.~\ifmmode~\check{S}\else \v{S}\fi{}imkovic, V.~Rodin, A.~Faessler, P.~Vogel,
  Phys. Rev. C \textbf{87}, 045501 (2013)

\bibitem{NME.Vaquero13}
N.L. Vaquero, T.R. Rodr\'{\i}guez, J.L. Egido, Phys. Rev. Lett. \textbf{111},
  142501 (2013)

\bibitem{NME.Barea15}
J.~Barea, J.~Kotila, F.~Iachello, Phys. Rev. C \textbf{91}, 034304 (2015)

\bibitem{NME.Pirinen15}
P.~Pirinen, J.~Suhonen, Phys. Rev. C \textbf{91}, 054309 (2015)

\bibitem{NME.Song17}
L.S. Song, J.M. Yao, P.~Ring, J.~Meng, Phys. Rev. C \textbf{95}, 024305 (2017)

\bibitem{NME.Simkovic18}
F.~\ifmmode~\check{S}\else \v{S}\fi{}imkovic, A.~Smetana, P.~Vogel, Phys. Rev.
  C \textbf{98}, 064325 (2018)

\bibitem{NME.Rath19}
P.K. Rath, R.~Chandra, K.~Chaturvedi, P.K. Raina, Frontiers in Physics
  \textbf{7}, 64 (2019)

\bibitem{Li.DM}
E.~Bertoldo, et~al., Journal of Low Temperature Physics \textbf{199}(1), 510
  (2020)

\bibitem{Li.Axion}
M.~Kr\ifmmode~\check{c}\else \v{c}\fi{}mar, Z.~Kre\ifmmode~\check{c}\else
  \v{c}\fi{}ak, A.~Ljubi\ifmmode \check{c}\else
  \v{c}\fi{}i\ifmmode~\acute{c}\else \'{c}\fi{}, M.~Stip\ifmmode \check{c}\else
  \v{c}\fi{}evi\ifmmode~\acute{c}\else \'{c}\fi{}, D.A. Bradley, Phys. Rev. D
  \textbf{64}, 115016 (2001)

\end{thebibliography}

\end{document}